\documentclass[a4paper,fleqn,usenatbib]{mnras}

\usepackage{newtxtext,newtxmath}

\usepackage[T1]{fontenc}
\usepackage{ae,aecompl}

\usepackage{graphicx}	
\usepackage{amsmath}	
\usepackage{amssymb}	

\title[Collision Between Molecular Clouds I]{Collision between Molecular Clouds I. The effect of the cloud virial ratio in head--on collisions}

\author[Tanvir et al.]{
Tabassum S Tanvir,$^{1,2}$\thanks{E-mail: tabassum.tanvir@anu.edu.au or tabassumtanvir61@gmail.com}
James E Dale,$^{2}$
\\
$^{1}$Research School of Astronomy and Astrophysics, The Australian National University, Cotter Rd, Weston Creek , ACT 2611, Australia\\
$^{2}$Centre for Astrophysics Research, University of Hertfordshire, College Lane, AL10 9AB, Hatfield, UK\\
}

\date{Accepted XXX. Received YYY; in original form ZZZ}

\pubyear{2018}

\begin{document}
\label{firstpage}
\pagerange{\pageref{firstpage}--\pageref{lastpage}}
\maketitle

\begin{abstract}
In a series of papers we investigate the effect of collisions between turbulent molecular clouds on their structure, evolution and star formation activity. In this paper we look into the role of the clouds' initial virial ratios. Three different scenarios were examined: both clouds initially bound, one cloud bound and one unbound, and both clouds initially unbound. Models in which one or both clouds are bound generate filamentary structures aligned along the collision axis and discernible in position--position and position--velocity space. If neither cloud is bound, no filaments result. Unlike in previous simulations of collisions between smooth clouds, owing to the substructure created in the clouds by turbulence before the collisions, dissipation of kinetic energy by the collision is very inefficient and in none of our simulations is sufficient bulk kinetic energy lost to render the clouds bound. Simulations where both clouds are bound created twice as much stellar mass than the bound--unbound model, and both these scenarios produced much more stellar mass than the simulation in which both clouds are unbound. Each simulation was also compared with a control run in which the clouds do not collide. We find the bound--bound collision increases the overall star formation efficiency by a factor of approximately two relative to the control, but that the bound--unbound collision produces a much smaller increase, and the collision has very little effect on the unbound--unbound cloud collision.\\
\end{abstract}



\section{Introduction}
Giant molecular clouds (GMCs) are the sites of star formation in galaxies, and interactions between these clouds are thought to be one of the major contributors towards their evolution and to the global regulation of star formation, with such events varying between tidal interactions to head--on collisions.\\
\indent Collisions between GMCs have been studied using hydrodynamical simulations on both smaller individual--cloud scales and global, galaxy--scale simulations. While the smaller--scale simulations have the resolution required to model individual collapsing prestellar cores, they cannot provide any information on the collision rates or parameters such as typical relative velocities, which must be treated as initial conditions. Conversely, global simulations provide a rate as to which such interactions between cloud may occur and how often, but cannot resolve in detail the outcomes of encounters.\\
\indent From the global simulations it can be concluded that multiple collisions could occur during a cloud's lifetime. \citet{2009ApJ...700..358T} explored the evolution of GMCs via cloud--cloud  collisions in a marginally unstable galactic disk, finding that collision timescales were a small and approximately constant fraction of the local orbital time which lends support to the theory that if collisions triggered star formation \citep{2000ApJ...536..173T, 2014MNRAS.445L..65F} they may be able to explain the empirical relation between the gas surface density and surface star formation rate \citep{1998ApJ...498..541K}. In a subsequent paper \citet{2011ApJ...730...11T} found that cloud collisions acted to \textit{suppress} star formation, particularly in encounters involving clouds of similar mass, but the authors caution that these results may be influence by lack of resolution in their global simulations. Their inferred specific star formation rate was ten times higher than observed in comparable galaxies and they propose that this is due to the absence of localised feedback mechanisms.\\
\indent In simulations of collisions between clouds with masses of 417 and 1625\,$\rm M_{\odot}$ and diameters of 7 and 15 pc, \citet{2014ApJ...792...63T} concluded that star formation is encouraged by the collisions, and that the number of cores formed during the collisions is proportional to the relative velocity between the clouds, although the subsequent growth of the cores depend on their stay in the high density shocked front. \citet{2015MNRAS.453.2471B} and \citet{2017MNRAS.465.3483B} also explored cloud collisions between parsec scale clouds and a relative velocity between 2 and 4\,km\,s$^{-1}$. Both the studies came to the conclusion that for smaller clouds slower collisions provided a greater efficiency in forming \textit{massive} stars. \citet{2017ApJ...841...88W} also studied the collision between molecular clouds with the added support from magnetic fields and found that the star formation rates and efficiencies are higher by a factor of ten in collision simulations.\\
\indent Observations have shown that there is a possible relation between cloud--cloud collision and massive star formation \citep{2009ApJ...696L.115F, 2014ApJ...780...36F, 2016ApJ...820...26F,2010ApJ...709..975O, 2011ApJ...738...46T, 2015ApJ...806....7T, 2017ApJ...835..142T}. Two super star clusters Westerland and NGC 3603 \citep{2009ApJ...696L.115F, 2014ApJ...780...36F,2010ApJ...709..975O} and the Triffid Nebula \citep{2011ApJ...738...46T} show relative velocity between 10 and 20 km/s under molecular lines observations. Broad bridge features \citep{2015MNRAS.450...10H} in the position--velocity diagrams also suggests these observations are a product of cloud--cloud collisions. \citet{2015ApJ...806....7T} also found another evidence of cloud--cloud collision in the Spitzer bubble RCW 120.\\
\indent To what extent GMCs are gravitationally bound is still debated. Observation and global simulations suggest that the clouds are probably borderline bound with virial ratios of $\rm \alpha$ $\sim$ 1-2 \citep{2013ApJ...777..173B,2016MNRAS.458.2443P}. However, measuring the virial ratio of a cloud observationally has always been challenging, arising in part from the various assumptions that have to be made (the shape , mass distribution and GMCs being isolated structures) and the results are quite uncertain, with errors of a factor of two. In any case, there is a considerable spread in inferred cloud virial ratios over around 2 decades from 0.1 to 10. In this paper, we examine the effect of the cloud virial ratio on the outcomes of collisions between clouds.\\
\indent We present the results from a recent batch of simulations of collisions of turbulent clouds using the smooth particle hydrodynamics code GANDALF. We explored the role of the virial ratio in terms of influencing the star formation rate and dynamics of collision product. In a companion paper we explored the role of other initial conditions namely the relative velocity and impact parameter on star formation rate and the dynamics of collision product. We did not introduce feedback into the simulations at this stage. This will be further explored in subsequent papers.\\
\section{Collisions between molecular clouds}
Star formation predominantly takes place in Giant Molecular Clouds (GMC), the coldest and densest parts of the Interstellar Medium (ISM) in Milky Way and other galaxies. It is still not understood what factors control the conversion from gas to stars, but collisions have long been thought to play a major role in their evolution. For this to be so, the probability of a given cloud colliding with another one during its lifetime must be appreciable. The typical cloud lifetime (and indeed what controls this timescale) is still unclear, but observational estimates currently mainly populate the range of a few to a few tens of Myr \citep{2015ARA&A..53..583H}.\\
\indent Estimating the collision timescale observationally is exceedingly difficult, and instead we may turn to simulations of whole galactic disks. \cite{2015MNRAS.446.3608D} examined this quantity in some detail. They observed, as did \cite{1980ApJ...238..148B}, that taking the number density of clouds and velocity dispersions prevalent in the Milky Way disk leads to collision times of $\sim300$\,Myr, much too long in comparison to cloud lifetimes to be of any interest. However, since clouds are concentrated in the regions of the spiral arms, their actual number densities, particularly at the time when they form, is considerably higher. In good agreement with their simulations, \cite{2015MNRAS.446.3608D} estimate a collision timescale $t_{\rm coll}\leq20$\,Myr and a mean free path between collisions of $\lambda\leq70$\,pc. This timescale is short enough that collisions are likely to be dynamically interesting, and the mean--free--path is comparable to the scale--height of molecular gas in a typical spiral, so that GMC collisions is likely to be a three--dimensional, rather than a two--dimensional, problem. We use these numbers to inform our choice of collision parameters.\\
\indent Consider a population of GMCs moving through the ISM at random. The cloud mass function is given by a power law with index $\beta$. The collision timescale (the inverse of the collision rate) for objects of mass $M_{i}$ travelling at speed $v_{0}$ through a medium containing other objects of mass $M_{j}$ with number density $n_{0}$ is given by
\begin{equation}
\tau_{j}=\left\{ n_{0}\left(\frac{M_{j}}{M_{0}}\right)^{-\beta}\times\pi(R_{i}+R_{j})^{2}\times\left[1+\frac{2G(M_{i}+M_{j})}{(R_{i}+R_{j})v_{0}^{2}}\right]v_{0}\right\}^{-1}.
\end{equation}
\indent This expression can be simplified somewhat by asserting that clouds have a constant column density, so that $M_{i}=aR_{i}^{2}$, where $a$ is a constant with dimensions of surface mass density:
\begin{equation}
\tau_{j}=\left\{n_{0}\left(\frac{M_{j}}{M_{0}}\right)^{-\beta}\times\frac{\pi}{a}\left(M_{i}^{\frac{1}{2}}+M_{j}^{\frac{1}{2}}\right)^{2}\times\left[1+\frac{2Ga^{\frac{1}{2}}(M_{i}+M_{j})}{\left(M_{i}^{\frac{1}{2}}+M_{j}^{\frac{1}{2}}\right)v_{0}^{2}}\right]v_{0}\right\}^{-1}.
\end{equation}
\indent To compute what kind of collisions occur most often, the above expression can be multiplied by the number density of clouds of mass $M_{i}$, yielding a collision rate \textit{per unit volume}
\begin{equation}
\frac{q_{ij}}{V}=\left\{n_{0}^{2}\left(\frac{M_{i}M_{j}}{M_{0}^{2}}\right)^{-\beta}\times\frac{\pi}{a}\left(M_{i}^{\frac{1}{2}}+M_{j}^{\frac{1}{2}}\right)^{2}\times\left[1+\frac{2Ga^{\frac{1}{2}}(M_{i}+M_{j})}{\left(M_{i}^{\frac{1}{2}}+M_{j}^{\frac{1}{2}}\right)^{2}v_{0}^{2}}\right]v_{0}\right\}.
\end{equation}
\begin{figure}
\includegraphics[width=0.45\textwidth]{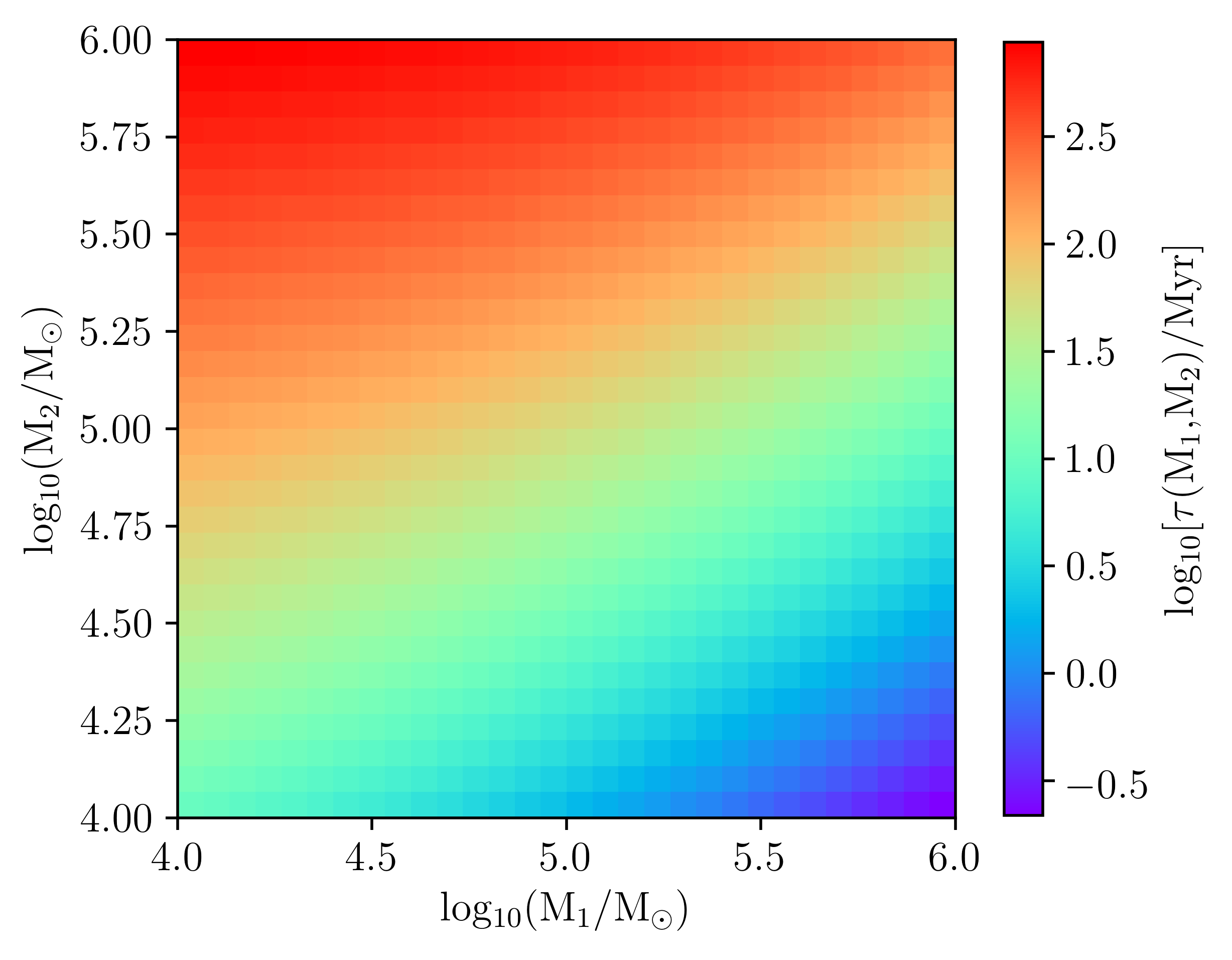}
\includegraphics[width=0.45\textwidth]{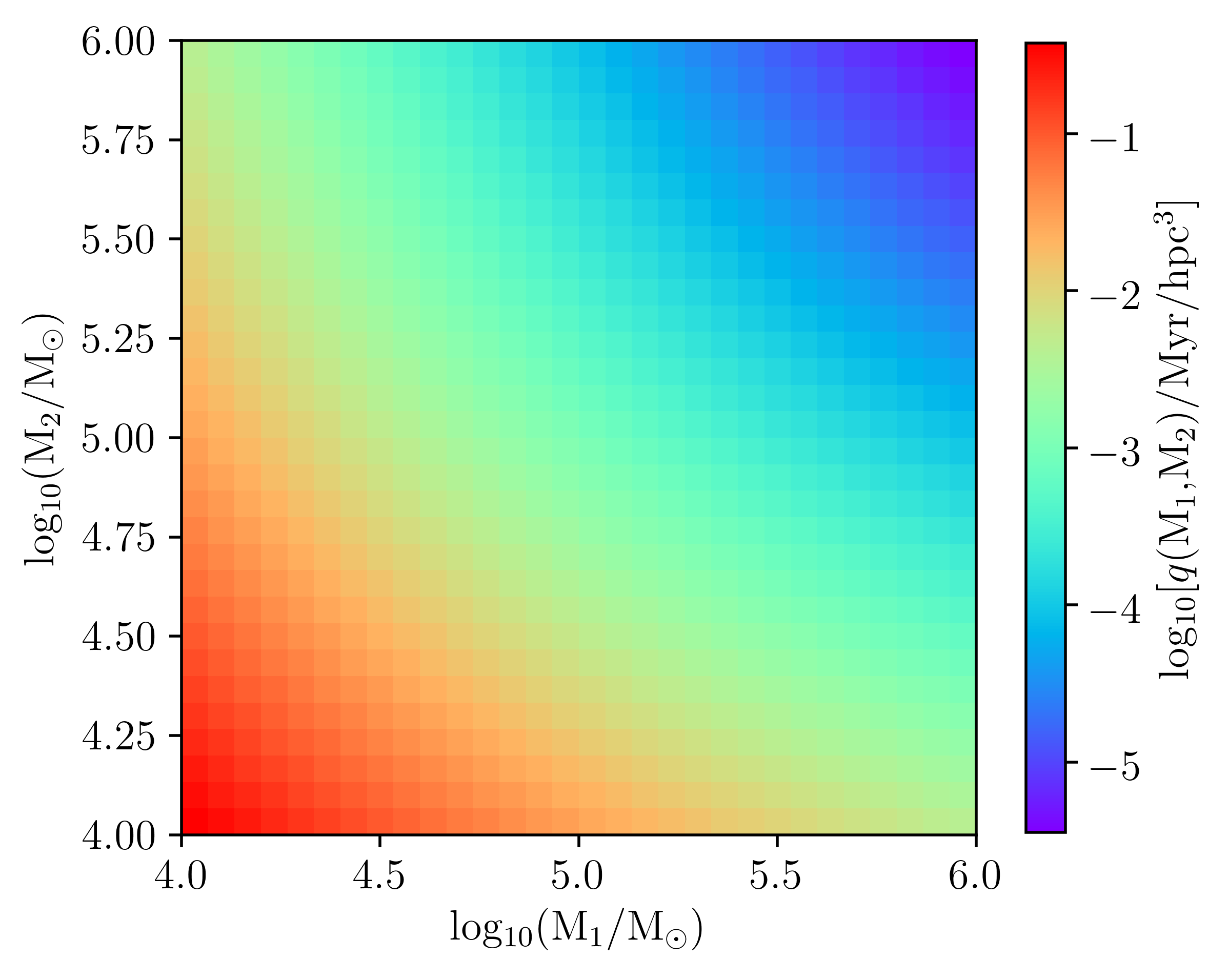}
\caption{Left panel: Given a cloud of mass M$_{1}$, the logarithm of the collision timescale in Myr with a cloud of mass M$_{2}$. Right panel: The logarithm of the rate per Myr per cubic hpc of collisions between clouds of mass M$_{1}$ and M$_{2}$.}
\label{fig:tau12}
\end{figure}
\indent Inserting reasonable values for the parameters in these two expressions allows the collision timescales and collision rates per unit volume to be visualised, shown in Figure 1. The mean molecular cloud mass is $\sim10^{5}$\,M$_{\odot}$, so we choose to examine masses ranging from one order of magnitude smaller to one order of magnitude higher. The number density $n_{0}$ we set to 1 per (65\,pc)$^{3}$ (a cube whose side length is close to the mean free path inferred by \cite{2015MNRAS.446.3608D}, and we choose $a=10^{-2}$\,g\,cm$^{-2}$ and $v_{0}=10$\,km\,s$^{-1}$. For the $10^{4}$\,M$_{\odot}$ clouds, this yields a collision timescale with each other (neglecting the gravitational focusing term) of 20\,Myr. In Figure \ref{fig:tau12}, we choose to express the rate per cubic hectoparsec because individual clouds fill many cubic parsecs, while the galactic scaleheight is much less than a kiloparsec.\\
\indent These expressions and plots immediately allow some useful conclusions to be drawn:\\
\\
(i) Any given cloud is most likely to collide with other clouds of low mass, and the most frequent kind of collision is a low--mass cloud striking another low-mass cloud, because the cloud mass function dictates that lower mass clouds are so much more common.\\
\\
(ii) Conversely, \textit{given a population of clouds that samples the cloud mass function well}, the clouds that suffer the most collisions during their lifetimes are the most massive ones, because they have the largest geometric cross sections and the largest degree of gravitational focusing. However, the vast majority of collisions experienced by high--mass clouds will be with clouds of much lower mass, because there are so many more such clouds.\\
\\
\indent In this work, we choose to concentrate on interactions of low--mass clouds with each other, since these are the most frequent events. We defer discussion of other encounters to later publications.\\
\indent Molecular clouds are complex objects and even simple models require several parameters to characterise them, e.g. mass, radius, virial ratio, temperature, etc. In combination with the possible parameters governing the collision, e.g. impact parameter, relative velocity, initial separation, these lead to a potentially enormous parameter space.\\
\indent In each of this series of papers, we therefore vary one or two possible parameters at a time to begin sketching out which are most important and what they are likely to govern. In a companion paper (Dale et al. 2019, in prep., hereafter Paper II), we examine the influence of the impact parameter and relative velocity, two of the most basic parameters of the \textit{collision}.\\  
\indent In the simulations presented here, we instead investigate the effect of a crucial parameter of the \textit{clouds}, namely the virial ratio. The main questions we aim to answer are\\
\\
(i) Do collisions result in greater star formation rates or efficiencies than would have obtained in the same clouds if they never collided? In particular, can gravitationally--unbound clouds, which would not normally be expected to support much star formation activity, be induced to form stars substantially faster or more efficiently than they would otherwise?\\
\\
(ii) Do the collisions result in the two clouds merging to form single objects?\\
\\
\indent There are several important timescales in this study, some of which are constant whilst others are variable. Individual clouds are characterised by their masses $M_{\rm cloud}$, radii $R_{\rm cloud}$ and turbulent velocity dispersion $\sigma$. Our model clouds are initially uniform with an imposed turbulent velocity field. The cloud parameters set the freefall time, which is constant in this study, since we confine ourselves to clouds of a single mass and initial radius. 
 \begin{equation}
 t_{\rm ff}= \sqrt{\frac{3\pi}{32G\rho}} \approx 5.2 {\rm\,Myr}
 \end{equation}
The cloud parameters also set the virial ratio, which in our study is permitted to vary. The virial ratio is defined by
\begin{equation}
\alpha = \frac{E_{\rm turb,0}}{\mid E_{\rm self,0} \mid}
\end{equation}
where the $E_{\rm turb,0}$ is the initial turbulent kinetic energy and $E_{\rm self,0}$ is the self gravitational potential energy of the clouds.
 \begin{equation}
E_{\rm turb,0}=\frac{M_{\rm cloud}\sigma ^{2}}{2}
\end{equation}
\begin{equation}
\mid E_{\rm self,0} \mid = \frac{GM_{\rm cloud}^{2}}{R_{\rm cloud}}
 \end{equation}
The velocity dispersion also sets the turbulent crossing/dissipation time $t_{\rm diss}$,
\begin{equation}
t_{\rm diss} \approx \frac{2R_{cloud}}{\sigma}
 \end{equation}
Collisions are characterised by three parameters: the impact parameter $b$, the initial relative velocity $v_{0}$ and the initial cloud separation $d$. These determine the collision time which, if $d>>b$ is given by 
\begin{equation}
t_{\rm coll} \approx \frac{d}{v_{0}}
 \end{equation}
Finally, the cloud crushing time scale is the time between the instant when  the clouds first touch and the time when as much of the clouds as is going to enter the shocked collision region has done so. The cloud crushing time  $t_{\rm crush}$ is
\begin{equation}
t_{\rm crush} \approx \frac{2R_{\rm cloud}}{v_{0}} \approx {\rm2.0\,Myr}
\end{equation}
The initial conditions of simulations run in this study are presented in table 1.\\
\indent The ratios of these timescales can be expected to have a strong influence on the outcomes of the simulations. If the collision time (and therefore the cloud crushing time) is short compared to the turbulent crossing and freefall times, bound clouds will have little opportunity to develop structure or collapse before striking each other, and before most of the gas has entered the shocked collision region. Conversely, unbound clouds will have expanded only a little before impact. In this case, the clouds would be relatively smooth at the collision, leading to an approximately planar shock in which most of any star formation that does occur is likely to happen.\\
\indent Conversely, a collision time comparable to or longer than the crossing and freefall times results in the bound clouds developing significant structure, and possibly initiating star formation \textit{before} the collision occurs. Instead, unbound clouds will have expanded significantly before the impact.\\
The temperature T = 30K was chosen as intermediate between the coldest, densest material at $\sim$ 10K and the outer regions of the clouds where the temperature could be closer to $\sim$ 100K.\\
The speed of sound at 30K for our molecular weight of 2.36 amu is 324 $m s^{-1}$. The collision Mach numbers are all 30.9. For the bound clouds, the turbulent Mach number is 5.84, and for the unbound clouds it is 13.1. There is no background medium present, so the clouds are evolving \textit{in vacuo}.\\
\color{black}
\indent One of our main interests is simply to gauge the effects of collisions on the clouds' star formation rates and efficiencies. To achieve this goal objectively, we perform control simulations in which the clouds are allowed to evolve in isolation.\\

\section{Numerical Methods and Initial Conditions}
Our simulations are performed with the Smoothed--Particle Hydrodynamics (SPH) code {\sc gandalf} \citep{2018MNRAS.473.1603H}. SPH is a simple and flexible approach to numerical fluid dynamics, automatically providing large dynamic ranges in density and spatial resolution and with very good conservation properties for linear and angular momentum. Its Lagrangian nature and ability to operate without boundary conditions also allow it to handle complex geometries and problems in which large fractions of the simulation domain are nearly or completely empty with ease. It is therefore the ideal method for studying collisions of turbulent molecular clouds. We do not include the effects of magnetic fields, or any form stellar feedback in these initial simulations. This allows us to run the simulations for a long duration -- nearly 10\,Myr -- and to concentrate on the interaction between the clouds' initial velocity fields and the bulk motion of the collisions.\\
\indent \textsc{gandalf} solves the fluid equations using the grad--h SPH formalism \citep[][]{2002MNRAS.333..649S,2004MNRAS.348..139P} and a leapfrog kick--drift--kick integrator.  Self--gravitational forces are computed using an octal tree. Regions of gas which become gravitationally unstable and collapse to high densities are replaced by sink particles, which are used to model star formation. We use a sink particle formation threshold density of 10$^{-17}$\,g\,cm$^{-3}$. There is no background medium.\\
\indent The gas thermodynamics is computed using a barotropic equation of state \citep[e.g.][]{2011A&A...529A..27H} with a critical density of 10$^{-16}$\,g\,cm$^{-3}$. Given the  threshold for sink particle formation, the simulations are thus effectively isothermal, with a gas temperature of 30\,K. The gas is assigned a mean molecular weight $\mu=2.35$. Artificial viscosity forces are computed according to the \cite{1997JCoPh.136..298M} scheme, with $\alpha=1$, $\beta=2$.\\
\indent Our clouds are modelled with $10^{6}$ SPH particles each, giving a mass resolution of 0.5-1\,$M_{\odot}$ (since particles have 50 neighbours), and all clouds are given the same divergence--free turbulent velocity field with a power spectrum obeying $P(k)\propto k^{-4}$ appropriate for supersonic turbulence. The root--mean--square velocity of the turbulence is treated as an adjustable parameter which we use to generate a required virial ratio. The turbulence is purely initial and is not artificially driven during the simulations.\\
\indent We model the collision of clouds with equal masses and initially equal radii and uniform densities. The collisions examined here are head on, so the impact parameter $b$ is zero. Clouds are placed with their centres of mass $d=$40\,pc apart on the $x$--axis, and bulk velocities in the $x$--direction of $\pm v_{0}/2$, where $v_{0}$ is 10\,km\,s$^{-1}$ in these models.\\
\indent The clouds are seeded with turbulent velocity fields whose RMS velocities are scaled in order to control their virial ratios.\\
\begin{table*}
\centering
\begin{tabular}{ccccccccccccccccc}
\hline \hline
Run  & Mass & Radius &Temperature&$\rm b_{0}$&$\rm v_{0}$&$\rm \alpha$&$\rm t_{ff}$&$\rm t_{cross}$& $\rm \rho$ \\ 
&$\rm M_{\odot}$&pc&K&&\,km\,s$^{-1}$&&Myr&Myr& g$cm^{-3}$\\
\hline \hline
 A & 10000&10&30&0&10&1.0&5.2&8.61& $\rm 1.62\times 10^{-22}$\\
  B & 10000&10&30&0&10&1.0, 5.0&5.2&8.61,3.85& $\rm 1.62\times 10^{-22}$\\
    C & 10000&10&30&0&10&5.0&5.2&3.85& $\rm 1.62\times 10^{-22}$\\
\hline \hline
\end{tabular}
\caption{Simulation parameters: Run name, masses of individual clouds, radius of clouds, gas temperature, impact parameter, initial relative velocity, initial virial parameter(s), initial freefall time and initial crossing time and initial density}
\end{table*}
\section{Results}
This section is divided into three subsections. The first describes the morphology of gas and stars resulting from the collisions. In the second sub--section, we examine the star formation efficiencies of the simulations, and in the third, we discuss the question of whether the clouds become bound or not.\color{black} \\
\subsection{Morphology of the Collision Simulations} 
\begin{figure}
\graphicspath{{Figures/}}
	\centering
	 \includegraphics[width=0.8\columnwidth]{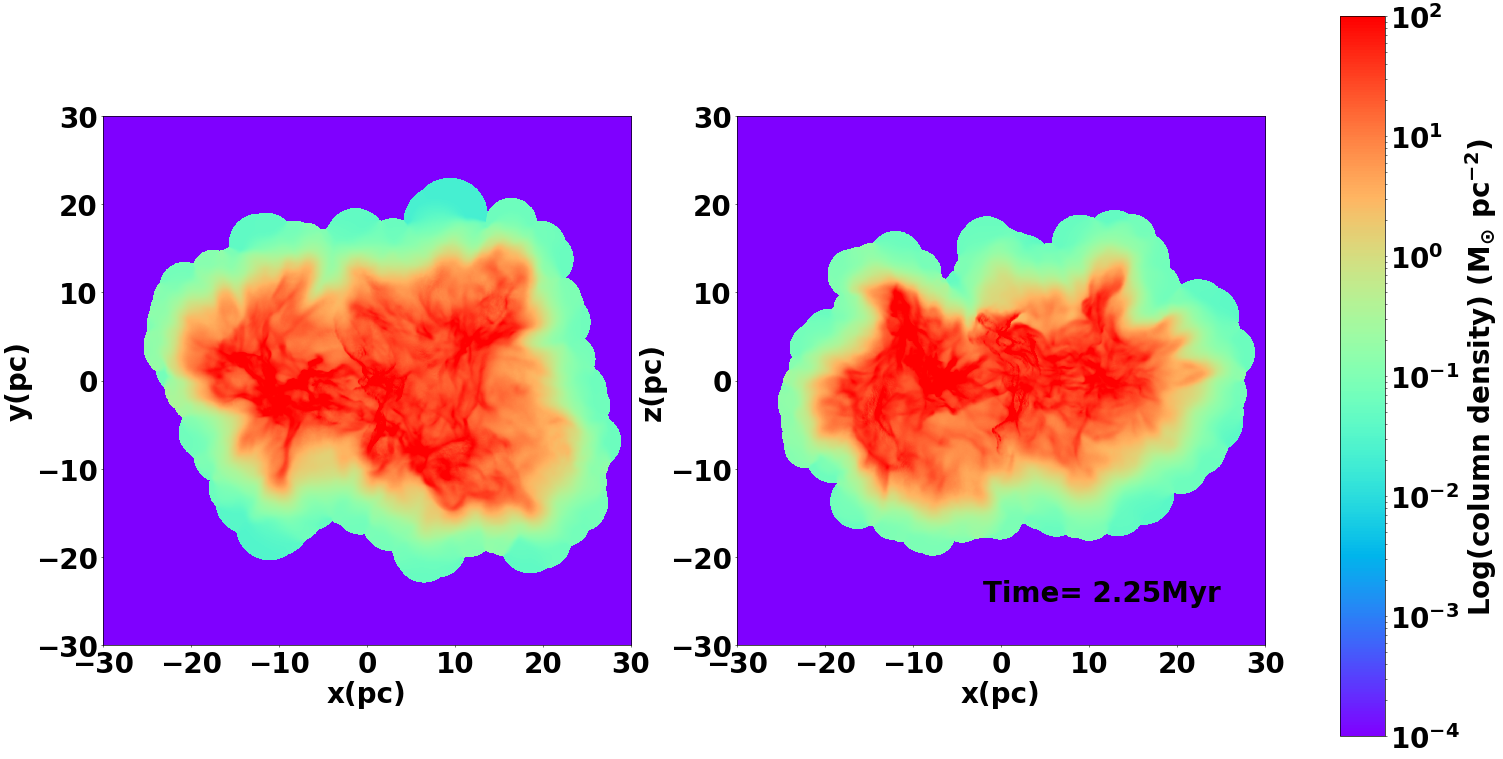}
      \includegraphics[width=0.8\columnwidth]{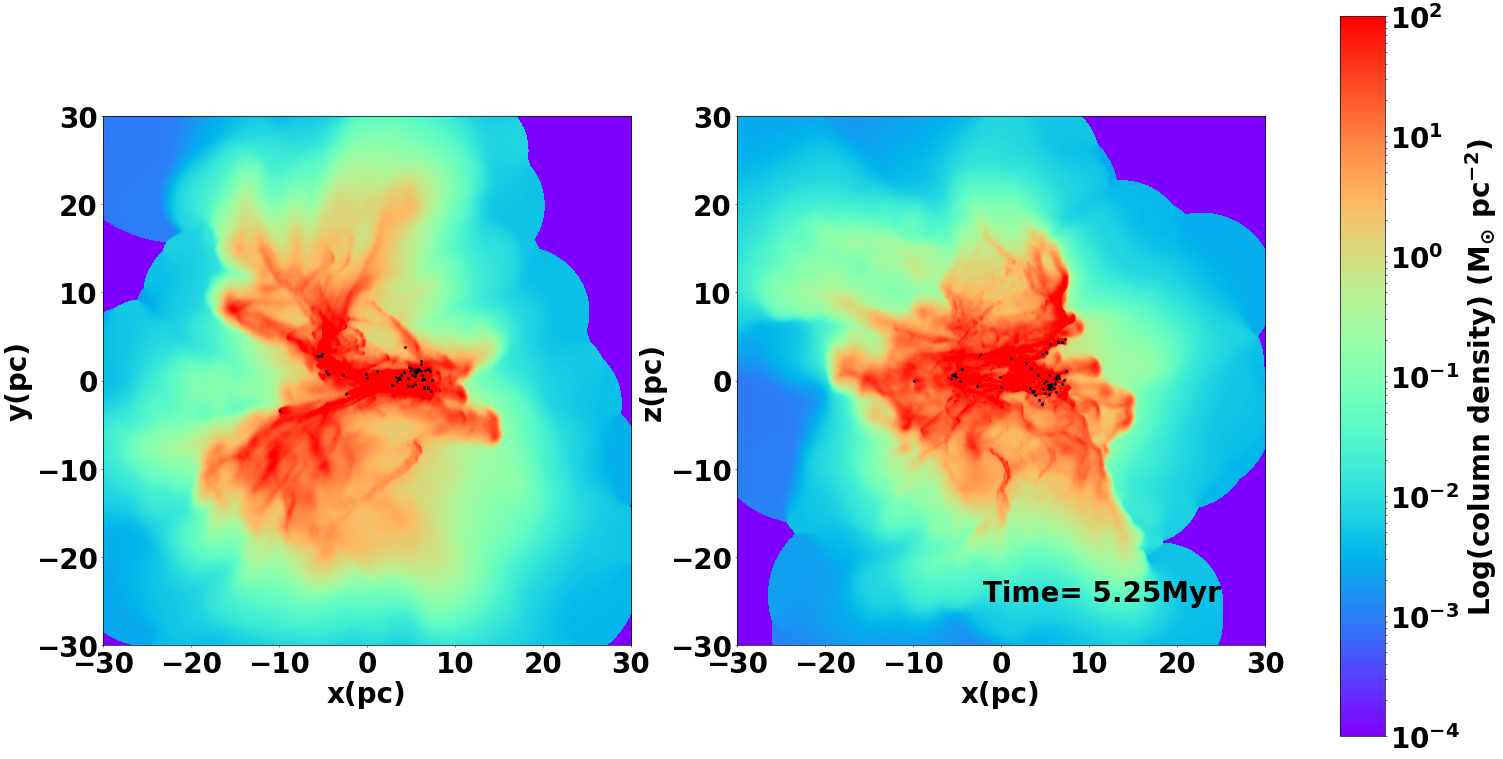}
            \includegraphics[width=0.8\columnwidth]{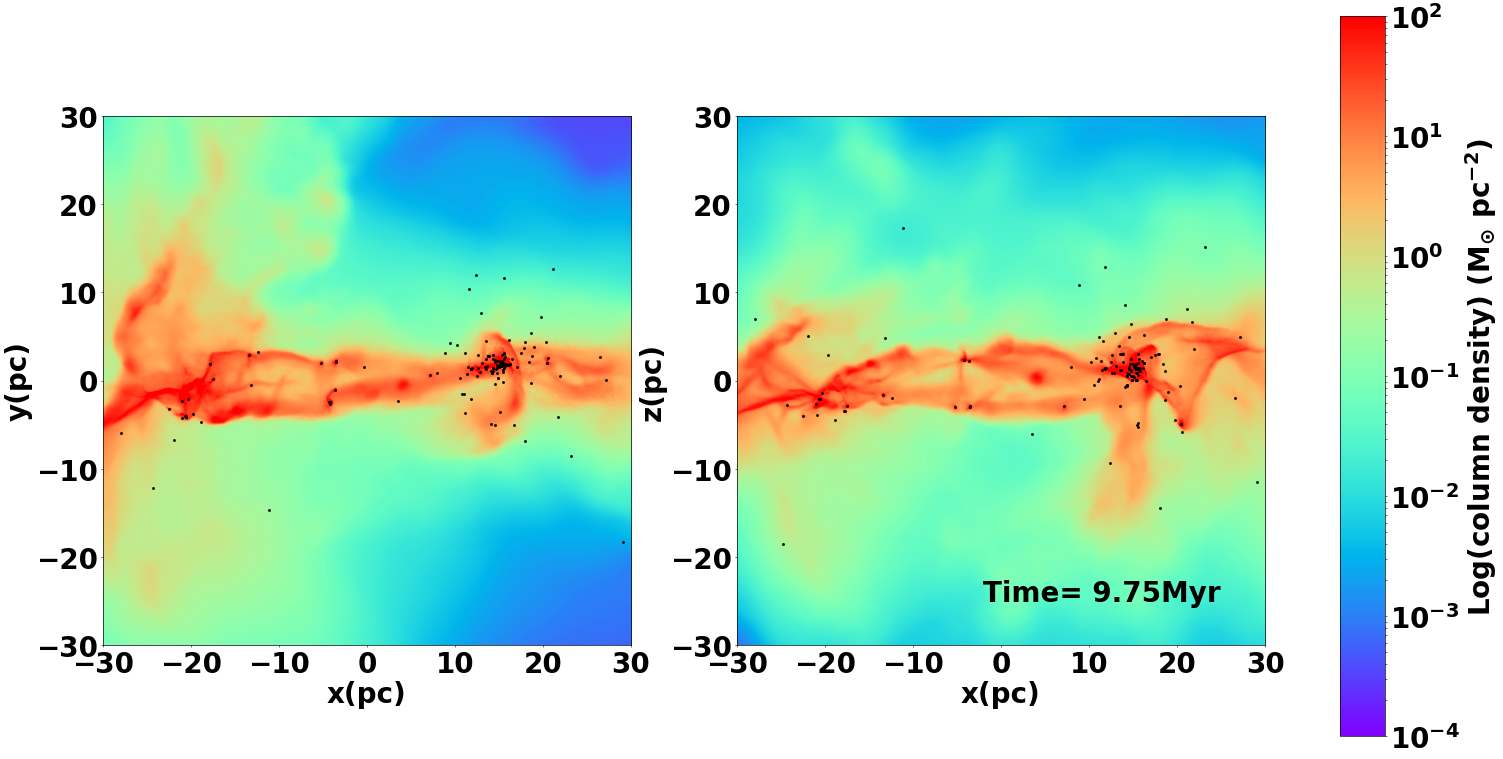}
  \caption[column-density]{Column density plots of from the Run A simulation snapshots. The first plot shows the clouds shortly after contact at a time of 2.25\,Myr. The second plot shows the collision at an intermediate time of 5.25\,Myr when the clouds are difficult to distinguish. The final plot shows the simulation at a late stage of 9.75\,Myr, when the remains of the original clouds are clearly separated.}
	\label{fig:runa_snaps}
\end{figure}

\begin{figure}
    \centering
             \includegraphics[width=0.8\columnwidth]{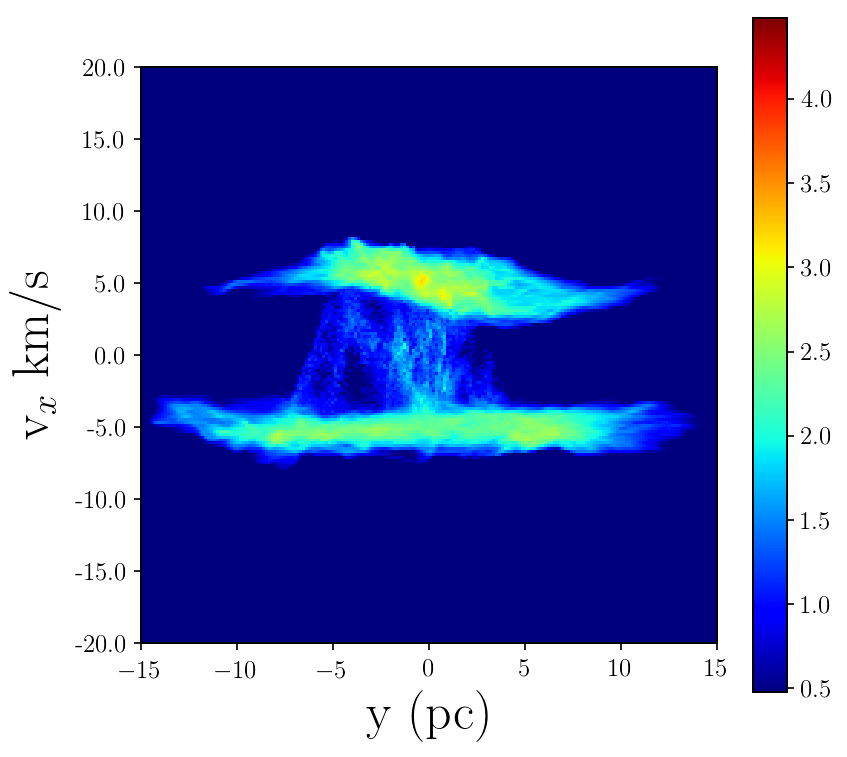}
                \includegraphics[width=0.8\columnwidth]{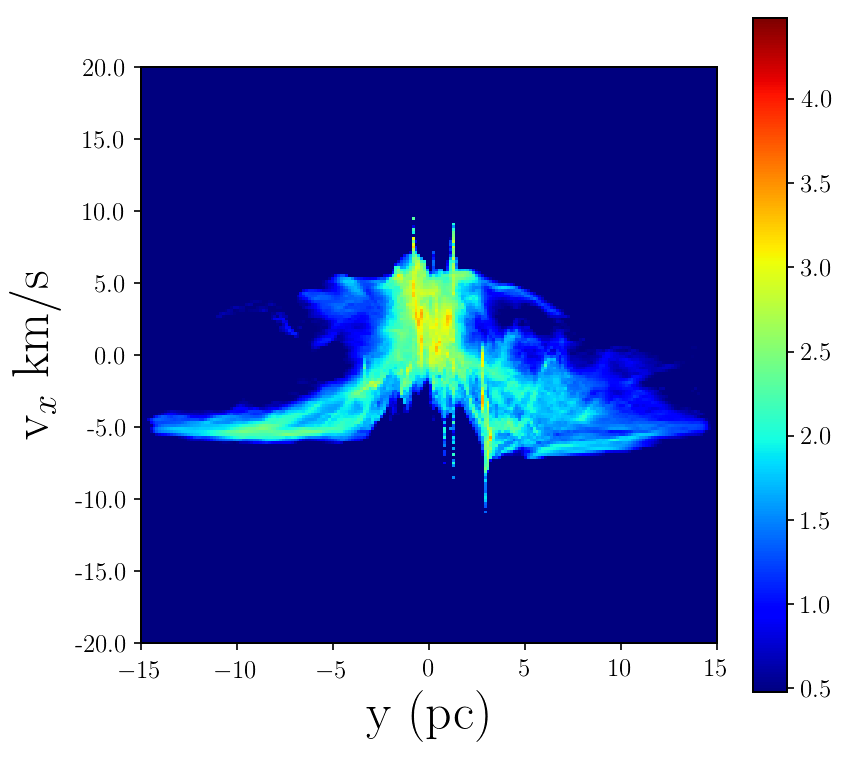}
                \includegraphics[width=0.8\columnwidth]{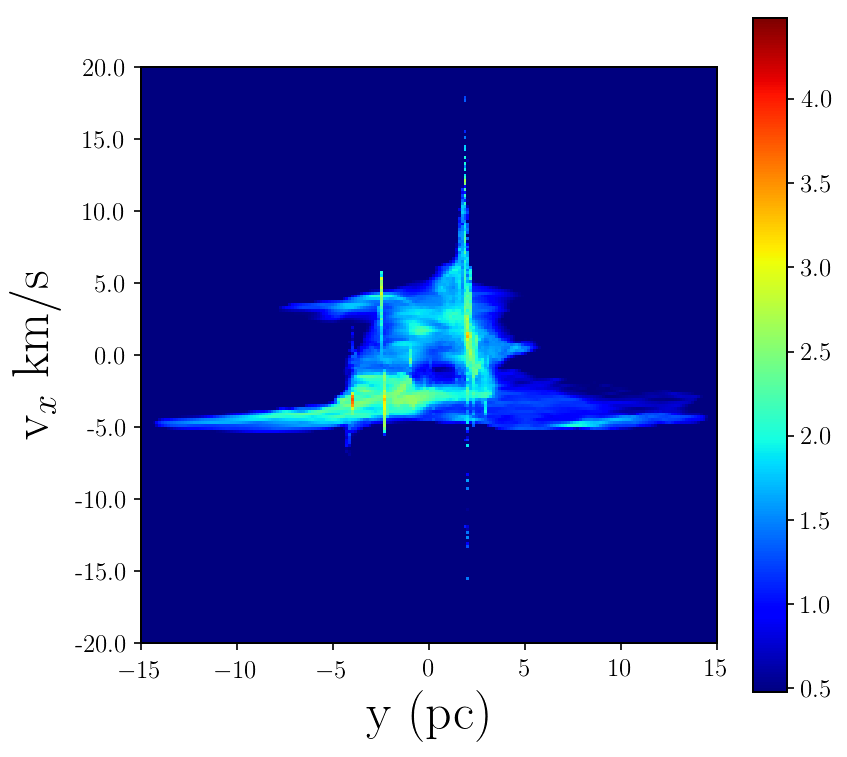}
    \caption{Position-velocity diagrams of Run A simulation. The upper panel is the position-velocity diagram of the clouds prior to the collision. The clouds are separated due to their pre collisional relative velocity. The second plot shows the state of the clouds during collision. The last plot is the position-velocity diagram of the clouds after the collision occurred. The timeline for this position-velocity diagrams are the same as the column density snapshots.}
    \label{fig:runa_pv}
\end{figure}

\textbf{Sim Run A}\\
In this run, the virial ratios of both clouds are 1.0, meaning that they are gravitationally bound. In this run the collision between the clouds occurs after 1.96 Myr. Figure \ref{fig:runa_snaps} shows column-density images of the simulation at three epochs after the collision. The first plot shows the simulation $\approx0.3$\,Myr after impact, at around 2.25 Myr.\\
\indent Since the collision time is shorter than the freefall or crossing times, the turbulent velocity fields of the clouds have had little time to dissipate and, although considerable structure has been generated, no stars have yet formed, and the gas distribution is still relatively isotropic.\\
\indent The second plot shows the state of the simulation at 5.25 Myr, approximately one initial freefall time. Star formation is now well underway and a star cluster has been formed, while the gas has acquired a highly anisotropic distribution.\\
\indent The third column density plot is 7.5 Myr after the collision. The collision product has acquired a roughly filamentary shape, aligned with the collision axis. This can be understood qualitatively from the highly anisotropic nature of the velocity field of the admixture of gas from the two clouds. The turbulent velocity dispersions of the original clouds are approximately 2\,km\,s$^{-1}$, sufficient to place them in virial equilibrium, whereas the initial collision velocity is 10\,km\,s$^{-1}$. When the clouds merge and their combined masses occupy the same volume, the velocity dispersion perpendicular to the collision axis is then too small to support the composite object against gravity in these directions. Conversely, the effective velocity dispersion along the collision axis is much greater than required to support the object against gravity in this direction. The overall result is then that the cloud is stretched along the collision axis and contracts perpendicular to it, resulting in an elongated structure. This process is discussed in more detail in Paper II. The denser regions of both the clouds have survived the collision, and now form the ends of the filament and it can be seen from that star formation has not occurred uniformly along the dense filament formed by the collision but is instead largely confined to two star clusters, one at either end of the collision product. This is likely connected to the fact that the denser parts of the clouds at the time of the collision are those that are able to propagate furthest from the point of impact.\\
\indent Only a limited amount can be learned from position--position data, since they contain no dynamical information. We therefore turn to position--velocity images. Figure \ref{fig:runa_pv} shows position-velocity diagrams from Run A, where the velocity is that along the collision axis (the $x$--axis), and the position is that along the  $y$--axis. The first frame shows the state of the clouds shortly after the collision at a time of 2.25\,Myr. The bulk of the two clouds are separated from each other by their initial relative velocity of 10 \,km\,s$^{-1}$. Each of the clouds have a width along the velocity axis determined by their turbulent dispersion. Connecting the two clouds is a small amount of material forming a `broad--bridge' feature \citep{2015MNRAS.450...10H,2015MNRAS.454.1634H}. This is composed of gas which has decelerated from the rest frames of the clouds. Note that, although this is a head--on collision symmetrical about $y=0$, the clouds' turbulent velocity fields have distorted their shapes show that the first material to come into contact is not symmetrical about the $y=0$ axis.\\
\indent The second position-velocity plot shows the state of the simulation at 5.25\,Myr. It is still possible to discern significant quantities of material separated by the original relative velocity. However, a large fraction of the gas mass is now found at intermediate velocities in the bridge feature and there is prominent notch around $y=0,v_{x}=-5$ \,km\,s$^{-1}$ where material from the cloud travelling in the negative $x$--direction has been decelerated. Much of the clouds' combined mass has entered the shocked contact layer and is concentrated in the broad--bridge feature near the collision axis in space and near zero in velocity. The vertical spikes visible indicate the presence of very high--velocity accretion flows centred on one of the sink particles. These features would in reality probably be optically thick and thus unobservable, but they give a convenient indication of where in the PV diagram star formation is underway. These features show that much of the star formation activity is concentrated around the filament or bridge in space, but that significant star formation occurs at close to the clouds' original velocities. This confirms the conclusion drawn from the column--density images that the densest parts of the two clouds have largely survived the collision and been little decelerated.\\
\indent The last position-velocity diagram shows the simulation 7.5\,Myr after the collision. This plot shows that substantial parts of the individual original clouds have survived the collision and retained their identity and velocity. The notch visible in the second plot has been largely filled in by motion of material inwards towards the collision axis. The dense, star--forming regions are aligned roughly along the collision axis at $y=0$, along with the much weakened bridge feature.\\

We have added the column density and position velocity diagrams of the controlled runs of the same clouds. Figure 4 shows the both column density and position-velocity plots at 7.5 Myr. Unlike the collision model the controlled runs do not create filament shape between the two clouds. Two star clusters have been formed in the denser parts of both the clouds. The position-velocity plot tells us even less about the clouds. Since both the clouds occupy the same width along the velocity axis it is quite difficult to distinguish each cloud.\\
\begin{figure}
\graphicspath{{Figures/}}
	\centering
	 \includegraphics[width=0.8\columnwidth] {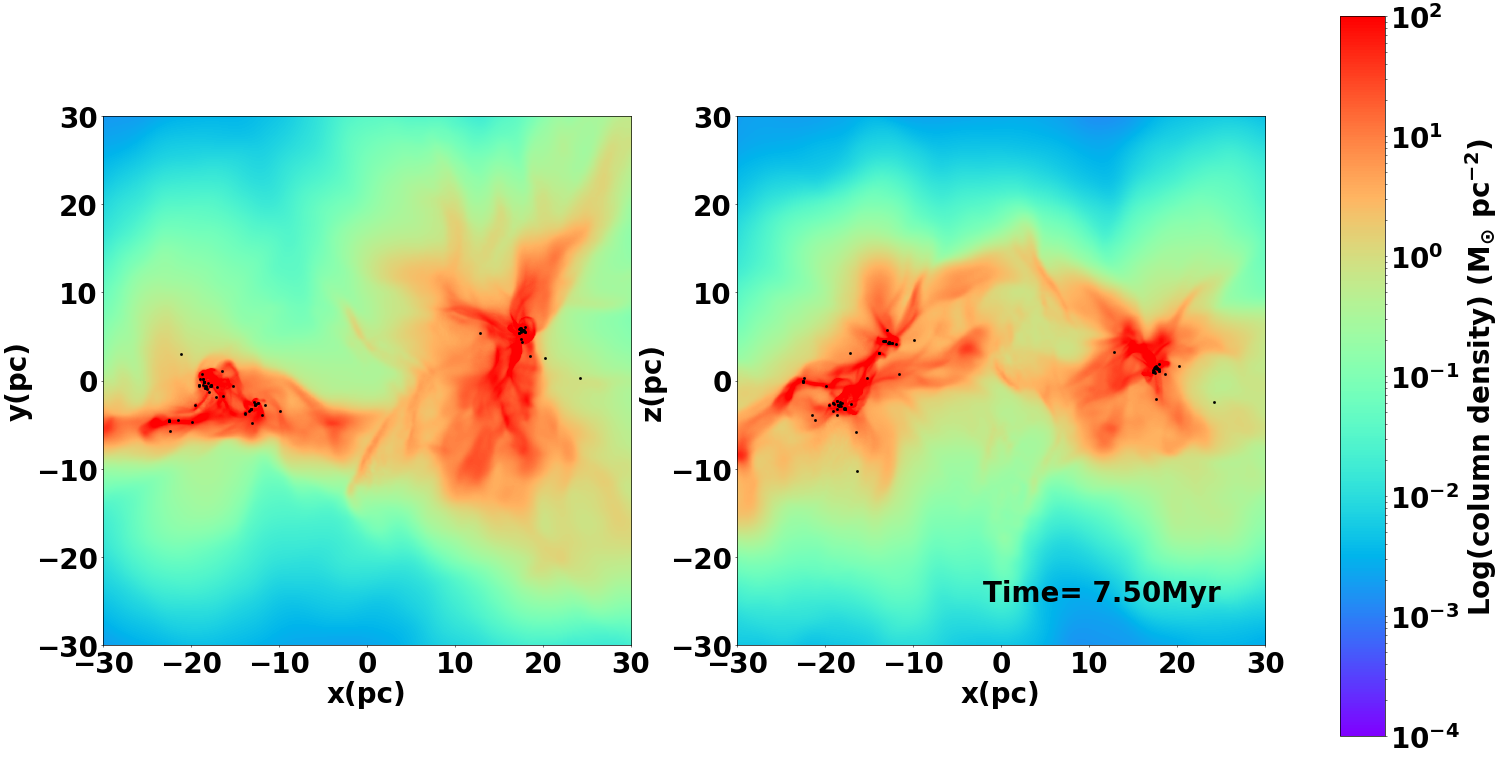}
      \includegraphics[width=0.8\columnwidth] {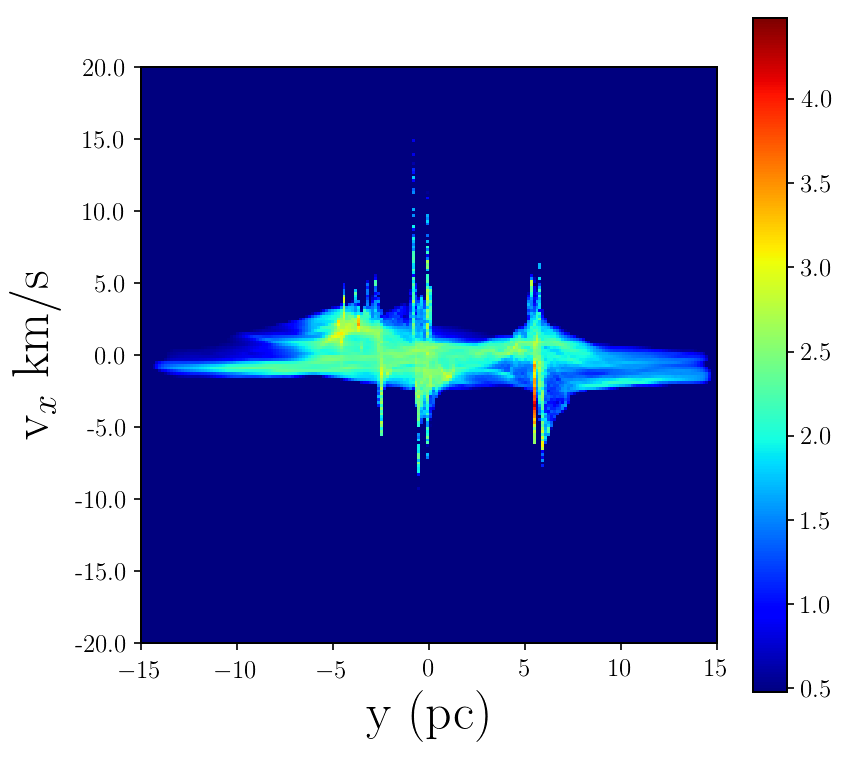}
           
  \caption[column-density]{Column density and position velocity plots of from the Run A simulation snapshots. The first plot shows the column-density at 7.5 Myr of the simulation. The second plot shows the position-velocity diagram of the same run at the same time. From the position-velocity diagram it is hard to distinguish the two clouds since both of them occupy the same space}
     \label{fig:runa_controlled}
	 \end{figure}
\color{black}
\\
\textbf{Sim Run B}\\
The virial ratios of the clouds in this simulation are 1.0 and 5.0, meaning that one of the clouds is substantially gravitationally unbound and therefore not expected to exhibit strong star formation.\\
\indent The cloud collision time is similar to Run A, occurring around 1.96 Myr. Figure 4 provides the column density plots of Run B. The unbound cloud begins at negative values of $x$ and is moving left--to--right. It is clear, by comparison with Figure \ref{fig:runa_snaps} that the unbound cloud has expanded and become substantially less dense than the bound cloud by the time of the collision, which accounts partly for the slower subsequent star formation in this cloud.\\
\indent In the second panel of Figure \ref{fig:runb_snaps}, it is scarcely possible to tell the two original clouds apart. However, the third panel, which shows the state of the simulation around 9.75 Myr, shows that the dense regions of the clouds again largely pass through each other, and the remnant of the bound cloud is now on the left side of the image. It can also be noticed that this simulation has produced more stars from the bound cloud (that on the left of the third panel) than the unbound. It is apparent that the two clouds have, to some extent, retained their identities despite the collision. The central filament is also visible as the concentration of material at a velocity and $y$--location close to zero.\\
  \begin{figure}
      \centering
      \includegraphics[width=0.8\columnwidth]{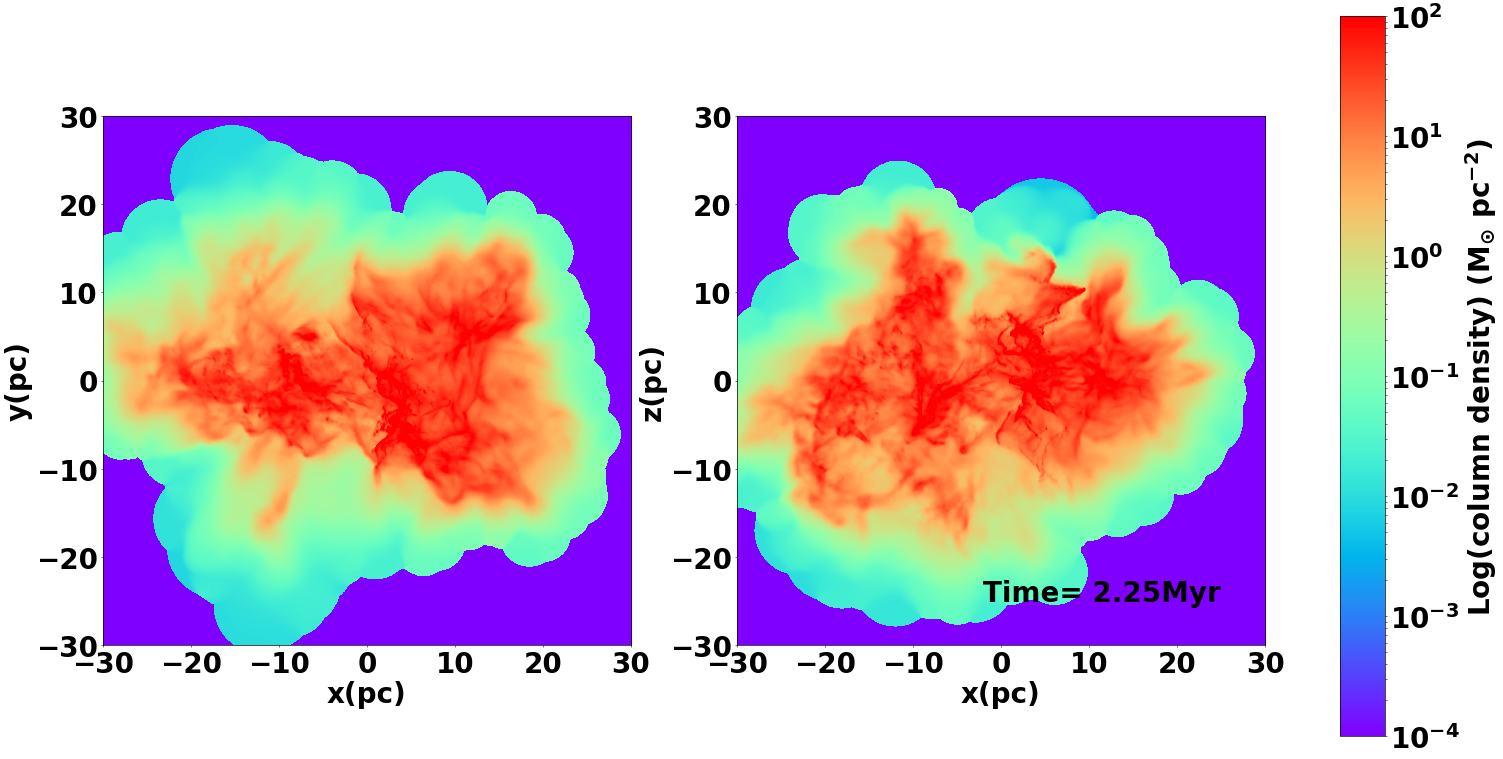}
      \includegraphics[width=0.8\columnwidth]{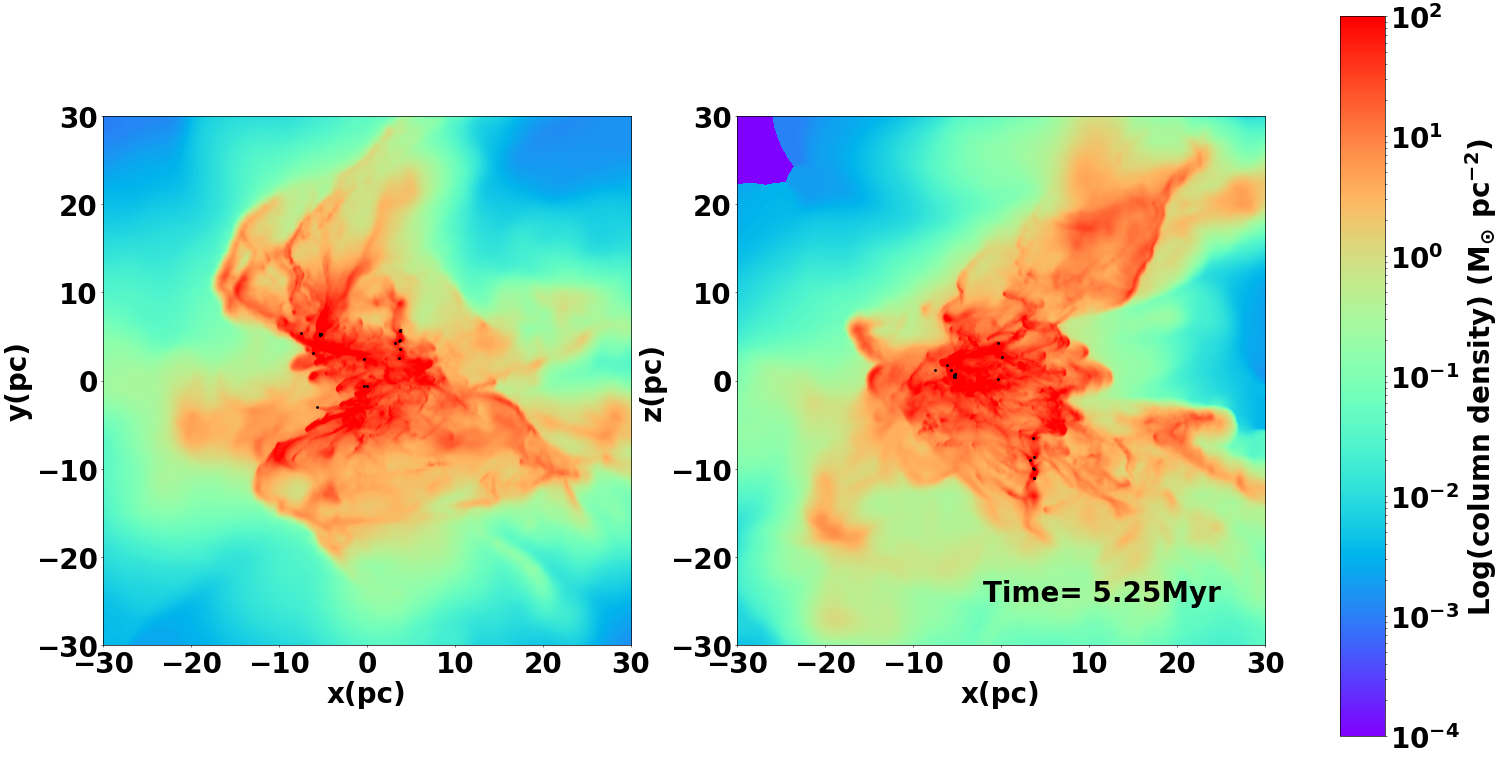}
      \includegraphics[width=0.8\columnwidth]{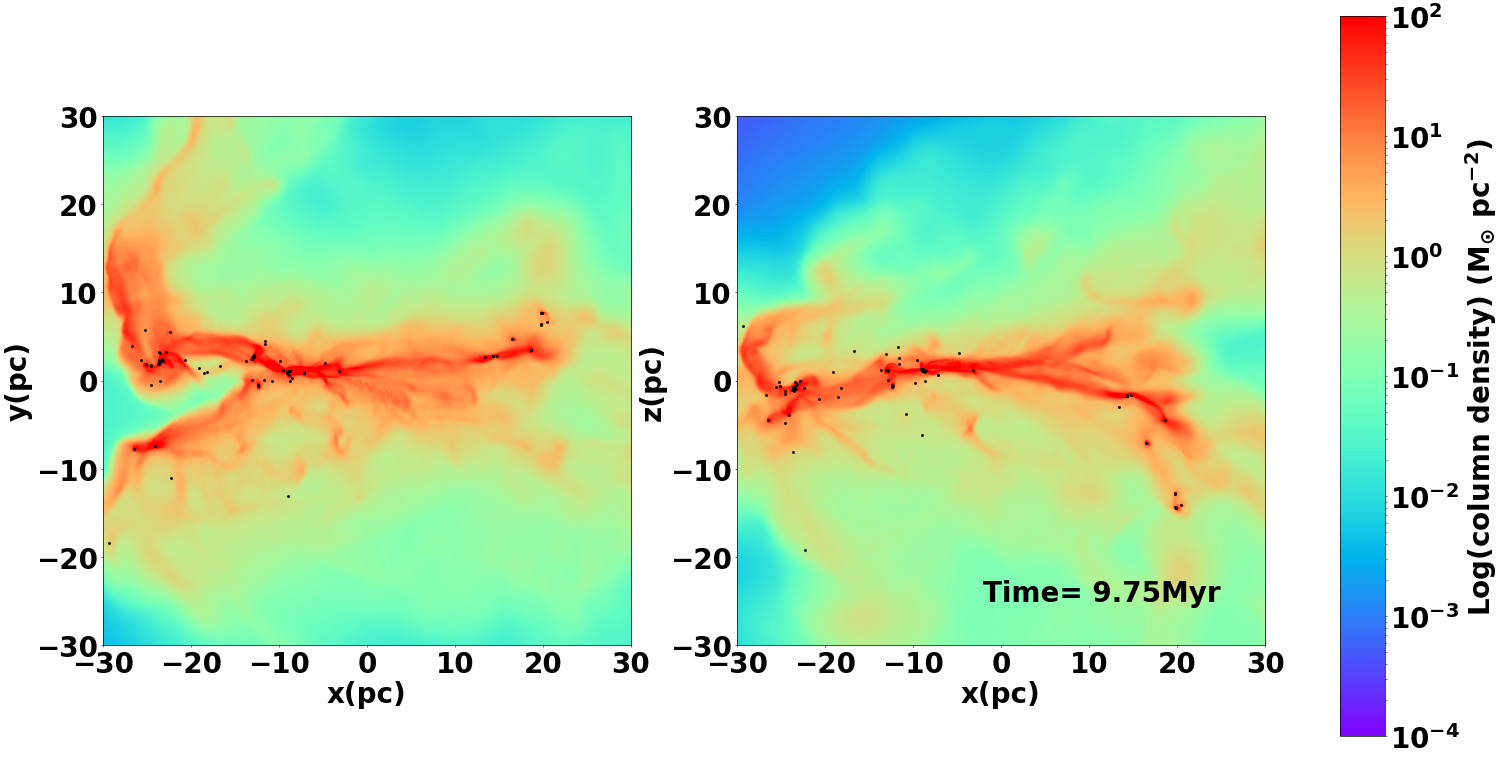}
      \caption[column-density]{Column density plots of from the Run B simulation snapshots. The first plot shows the clouds shortly after contact at a time of 2.25\,Myr. The second plot shows the collision at an intermediate time of 5.25\,Myr when the clouds are difficult to distinguish. The final plot shows the simulation at a late stage of 9.75\,Myr, when the remains of the original clouds are clearly separated.}
      \label{fig:runb_snaps}
  \end{figure} 
 \begin{figure}
      \centering
      \includegraphics[width=0.8\columnwidth]{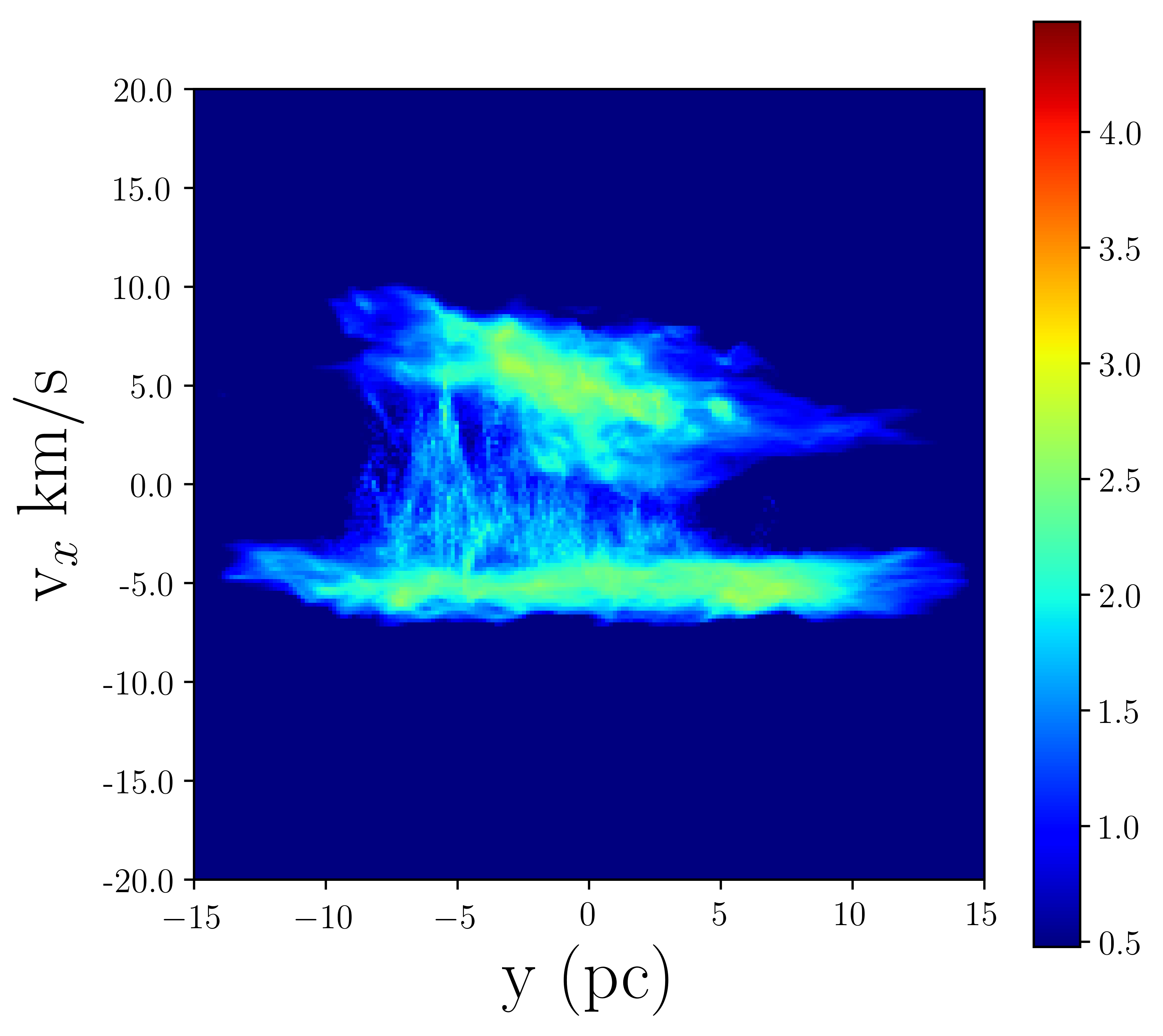}
      \includegraphics[width=0.8\columnwidth]{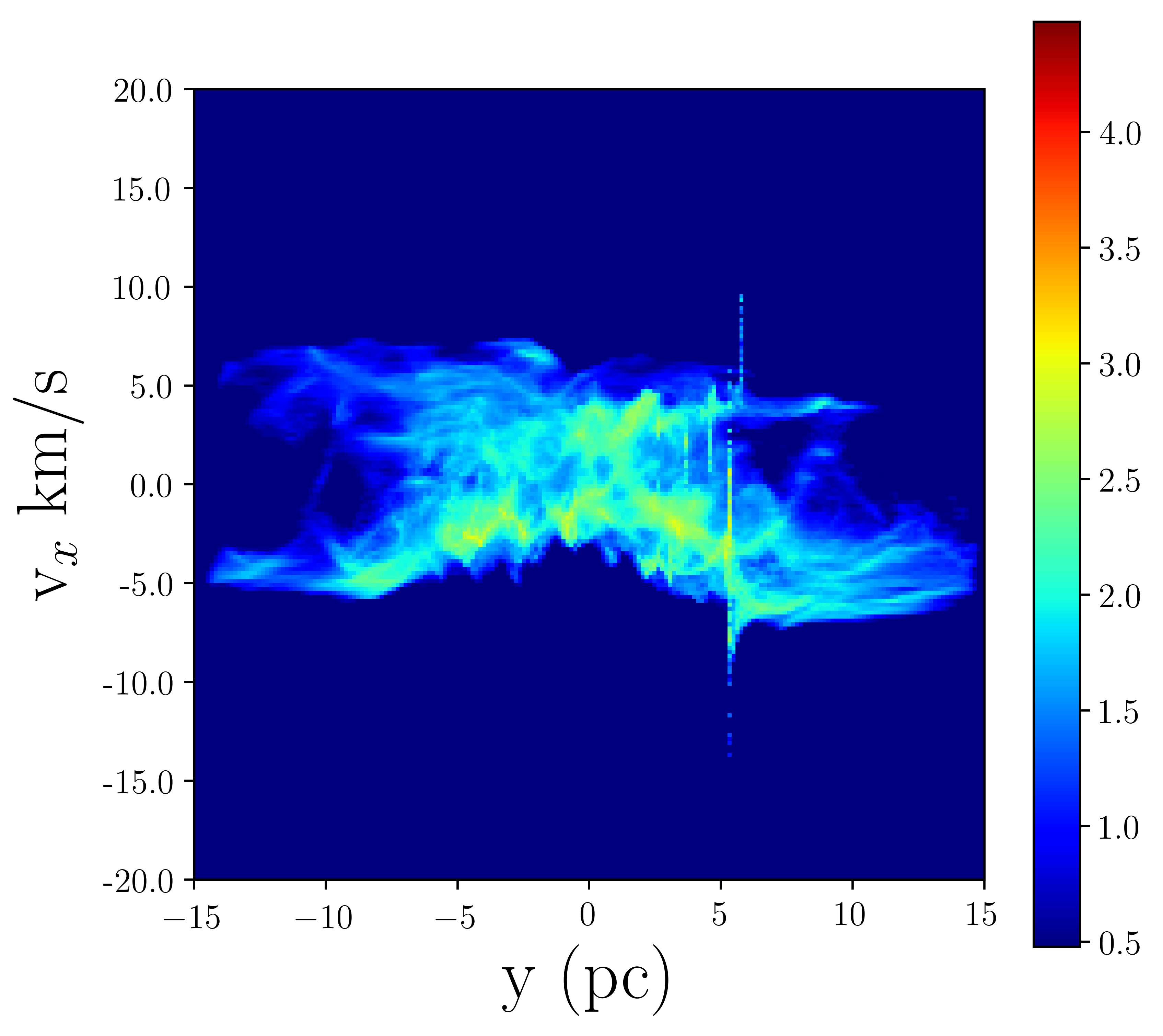}
      \includegraphics[width=0.8\columnwidth]{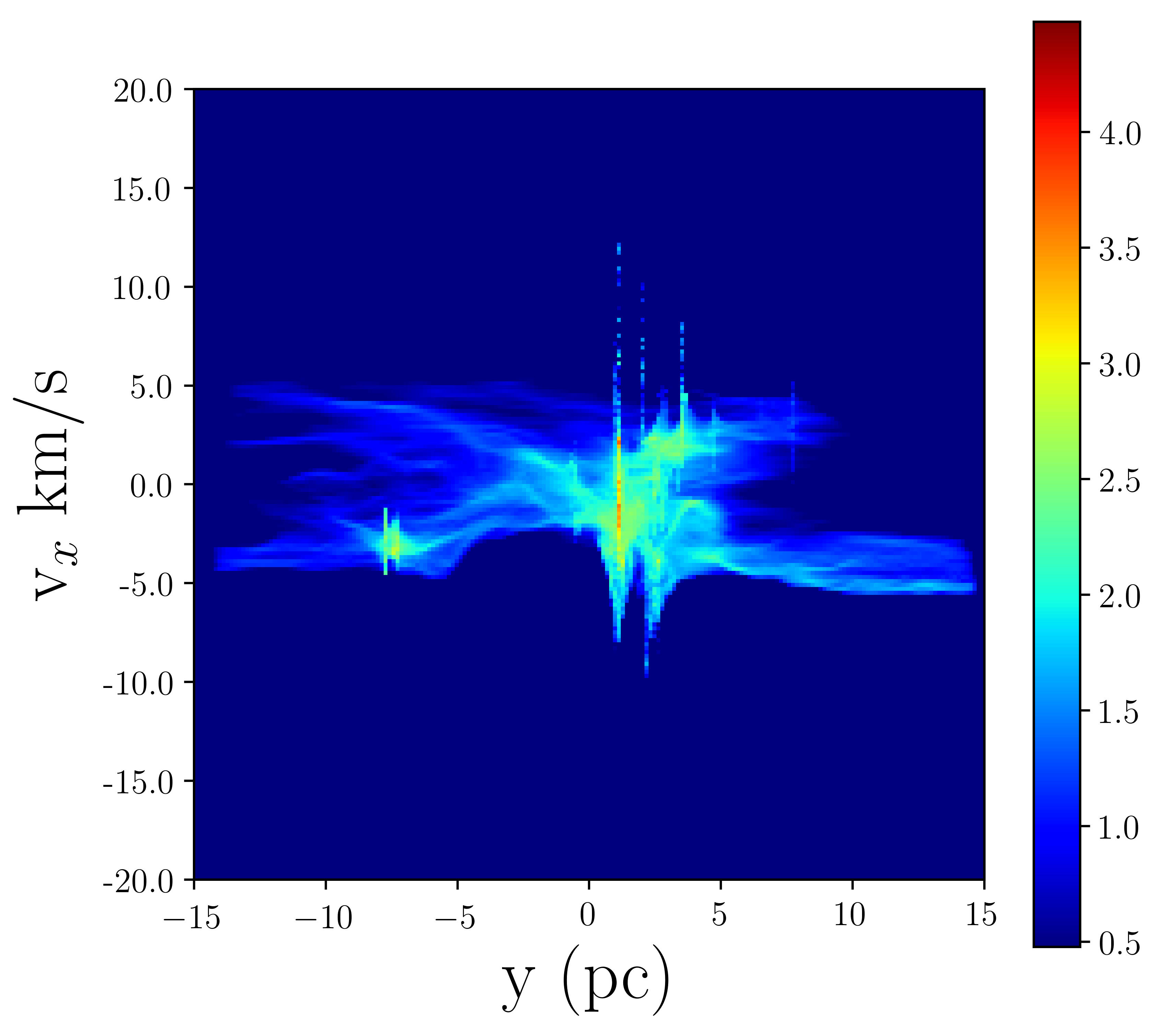}
      \caption{Position-velocity diagrams of Run B simulation. The upper panel is the position-velocity diagram of the clouds prior to the collision. The clouds are separated due to their pre collisional relative velocity. The second plot shows the state of the clouds during collision. The last plot is the position-velocity diagram of the clouds after the collision occurred. The timeline for this position-velocity diagrams are the same as the column density snapshots.}
      \label{fig:runb_pv}
  \end{figure}
\indent Figure \ref{fig:runb_pv} shows the position-velocity diagram of the simulation. The first plot shows the clouds being separated by their pre collisional velocity of 10\,km\,s$^{-1}$. The unbound cloud is evident by its thicker extent along the velocity axis of the plot.\\
\indent The second PV plot shows the clouds coming into contact with each other. It is much difficult to discern the clouds as separate objects than in the corresponding PV plot from Run A. In common with the Run A simulation, this simulation shows a bridge like feature between the clouds as they are starting to merge. However, the bridge is now so broad along the spatial dimension and of such complex structure that it is difficult to separate from the two clouds.\\
\indent The third panel of Figure 5 shows the position-velocity diagram 7.8\,Myr after the clouds have collided. The denser regions of the two clouds have survived the collision and are still discernible as separate features, with that originating from the bound cloud being somewhat stronger. The filament--like structure is clearly visible as a prominent feature in the PV plot with a velocity and $y$--location again close to zero. Accretion flows from the stars can be seen from the plots, and show that most of the star formation is associated either with the remains of the bound cloud, or with the filament. It is also noticeable that the cloud with the larger virial ratio is much more spread out in velocity compared to the more gravitationally bound object.\\

\begin{figure}
\graphicspath{{Figures/}}
	\centering
	 \includegraphics[width=0.8\columnwidth]{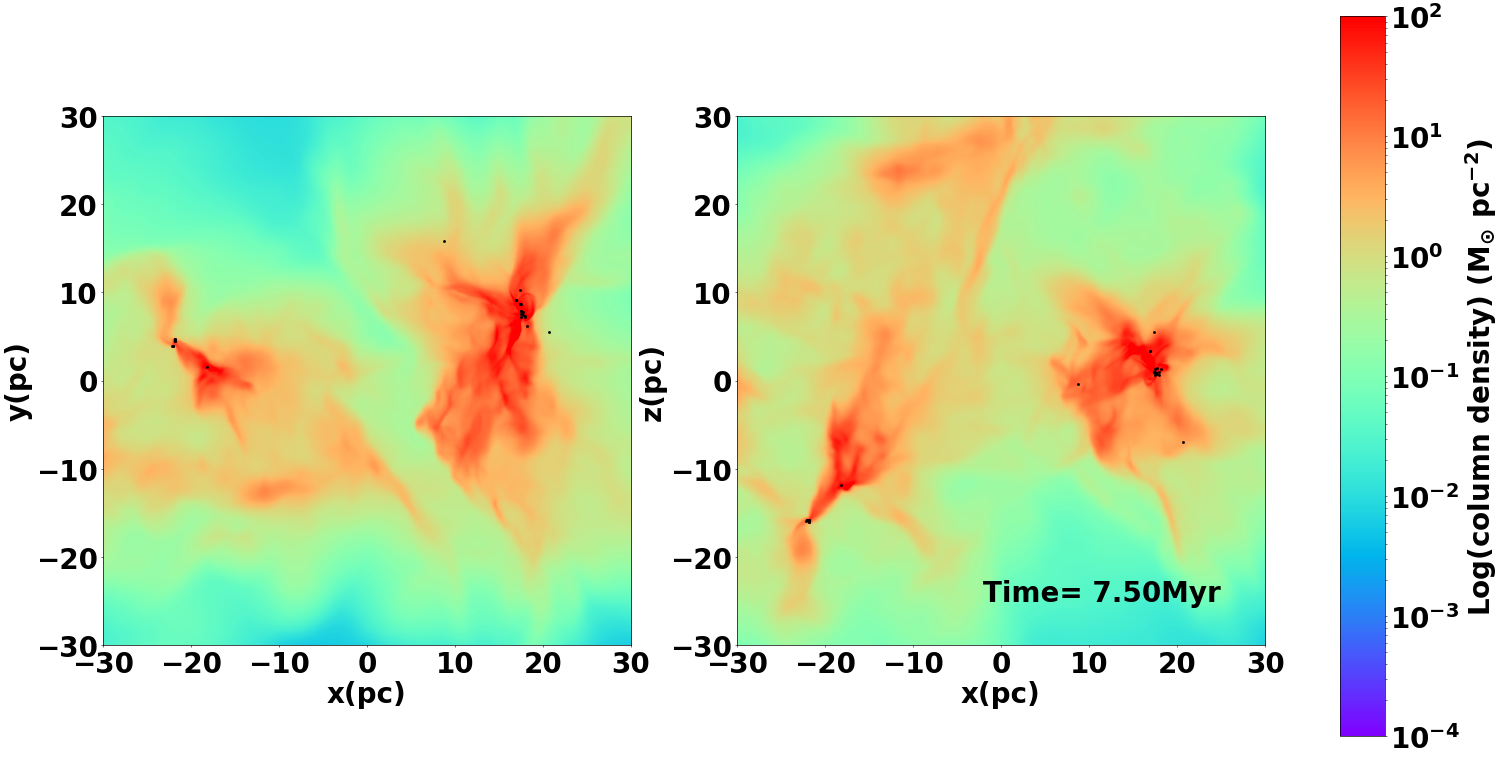}
      \includegraphics[width=0.8\columnwidth]{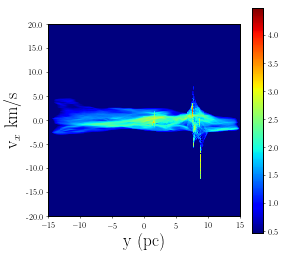}
           
  \caption[column-density]{Column density and position velocity plots of from the Run B simulation snapshots. The first plot shows the column-density at 7.5 Myr of the simulation. The second plot shows the position-velocity diagram of the same run at the same time. From the position-velocity diagram it is hard to distinguish the two clouds since both of them occupy the same space}
     \label{fig:runB_controlled}
	 \end{figure}

Column density and position velocity plots have been added from the controlled runs of the same simulation. Figure \ref{fig:runB_controlled} represents both the position-position and position-velocity plot of the controlled run at 7.5 Myr. The bound cloud is on the right side of the column-density plot and have formed more stars than the unbound cloud. Unlike the collision simulation a central filament structure is missing in this case. From the position-velocity diagram it is harder to determine each cloud separately since they both occupy the same width in the velocity axis. The vertical spikes present in the plot indicates towards the presence of very high-velocity accretion flow centred on one of the sink particles. \\
\color{black}

\begin{figure}
\graphicspath{{Figures/}}
	\centering
	 \includegraphics[width=0.8\columnwidth]{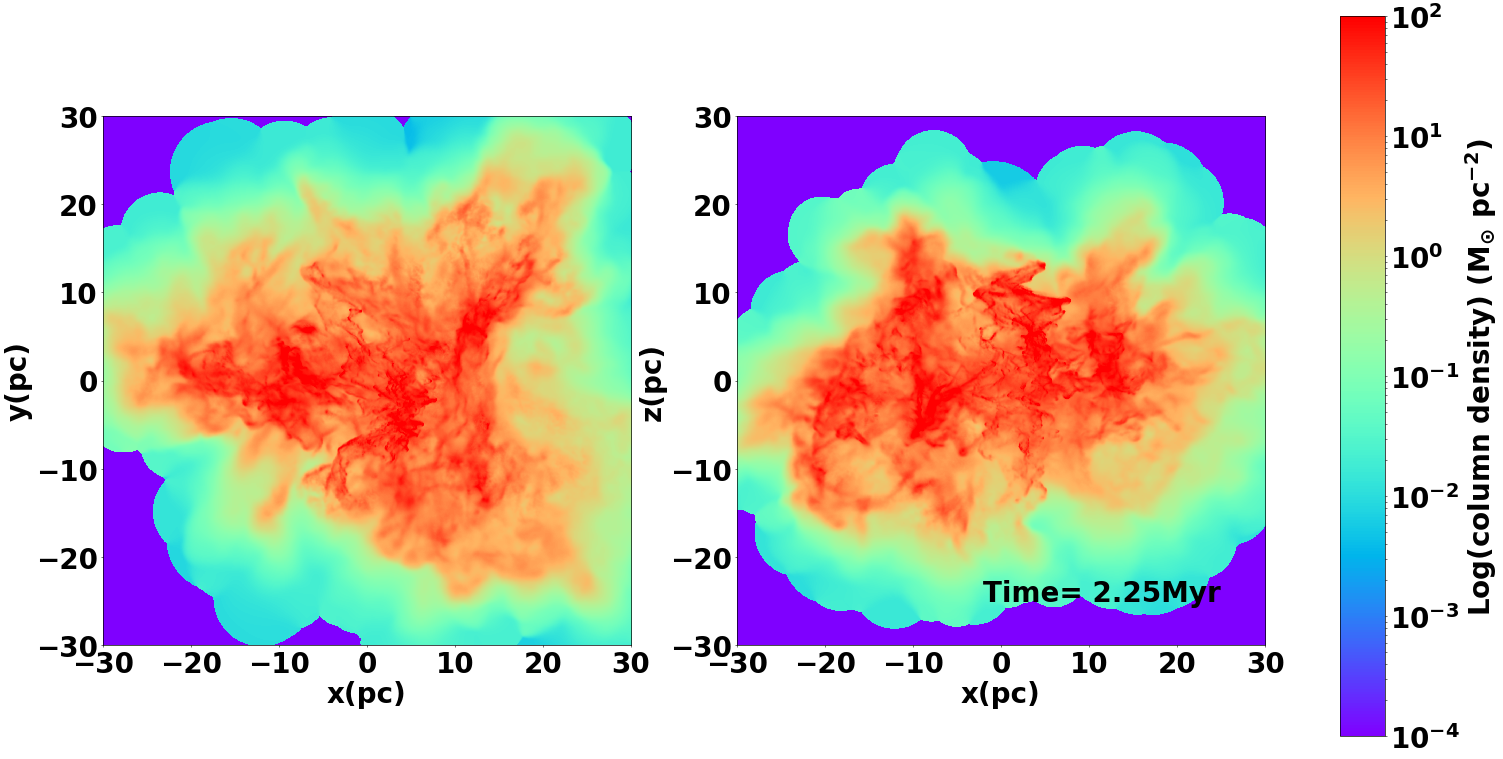}
      \includegraphics[width=0.8\columnwidth]{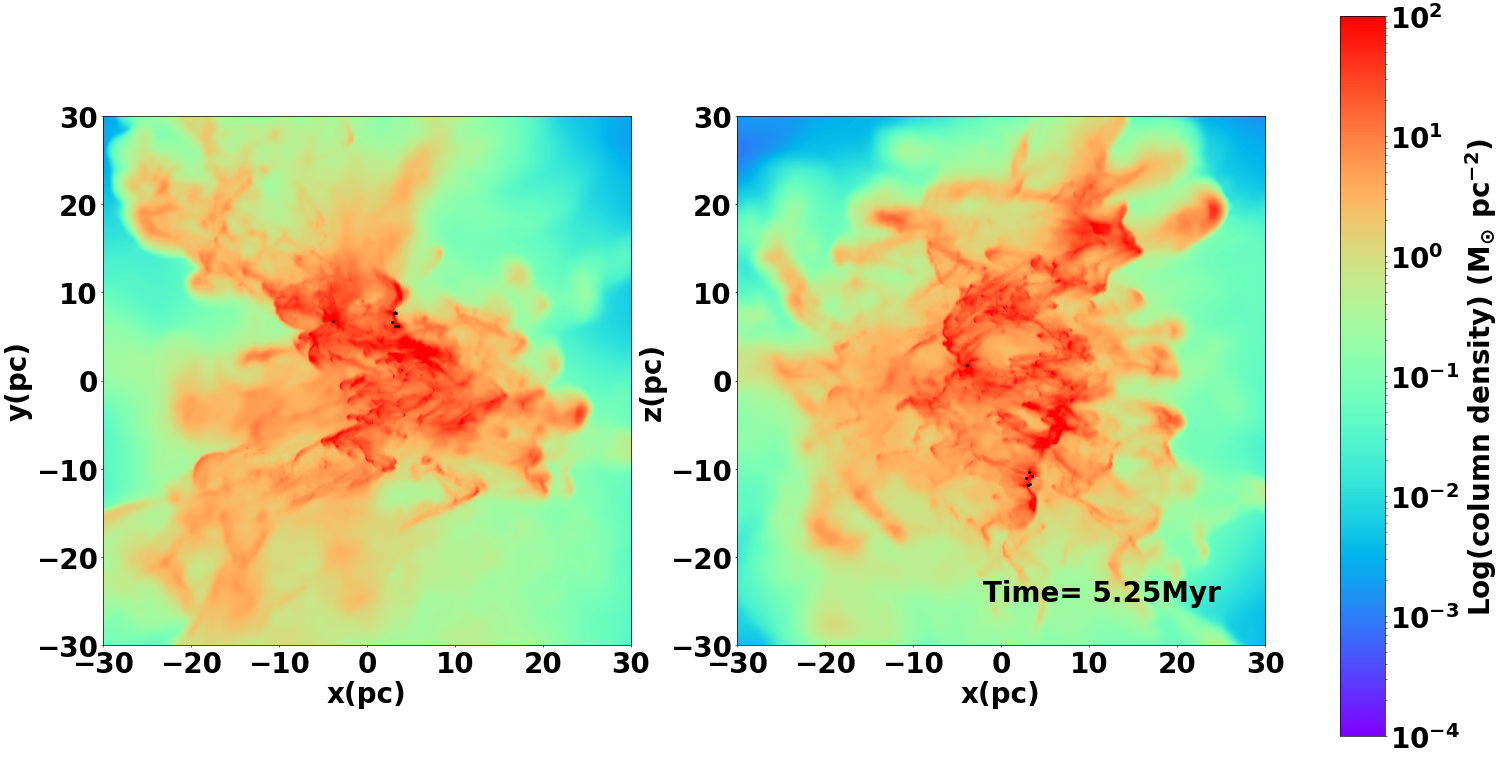}
            \includegraphics[width=0.8\columnwidth]{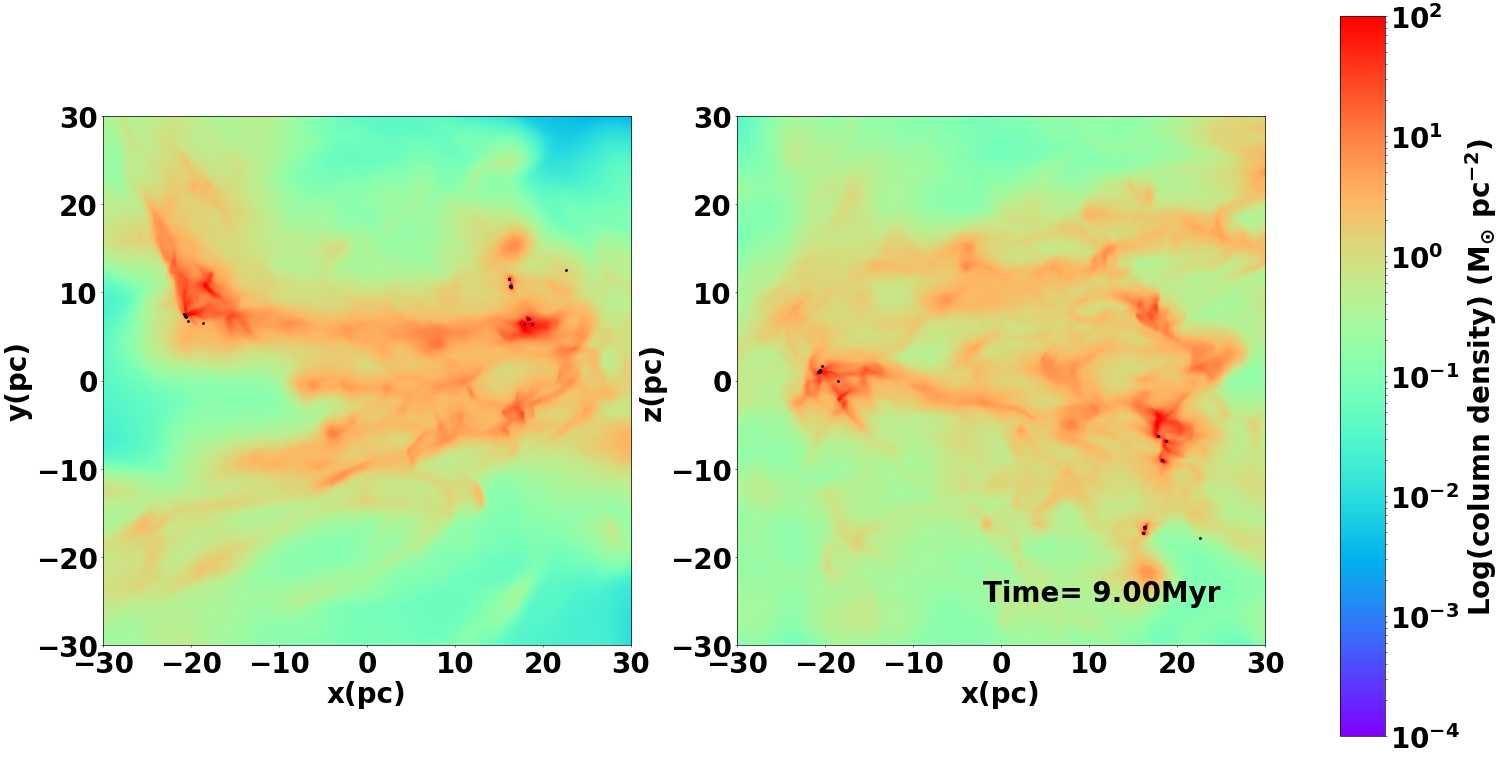}
  \caption[column-density]{Column density plots of from the Run C simulation snapshots. The first plot shows the clouds shortly after contact at a time of 2.25\,Myr. The second plot shows the collision at an intermediate time of 5.25\,Myr when the clouds are difficult to distinguish. The final plot shows the simulation at a late stage of 9.00\,Myr.}
	\label{fig:runc_snaps}
\end{figure}

\begin{figure}
\graphicspath{{Figures/}}
	\centering
	 \includegraphics[width=0.8\columnwidth]{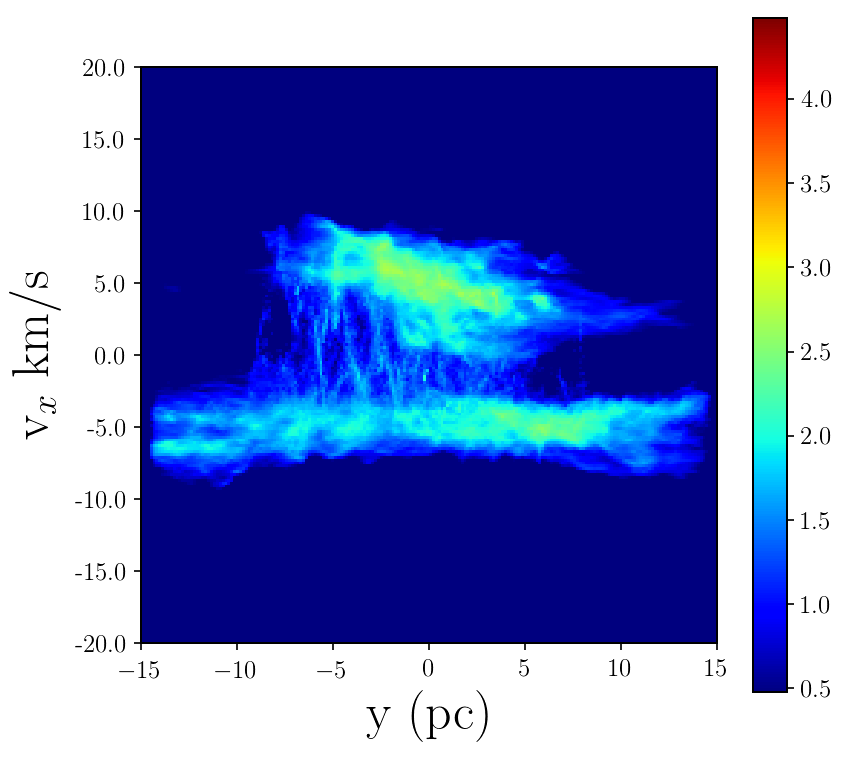}
      \includegraphics[width=0.8\columnwidth]{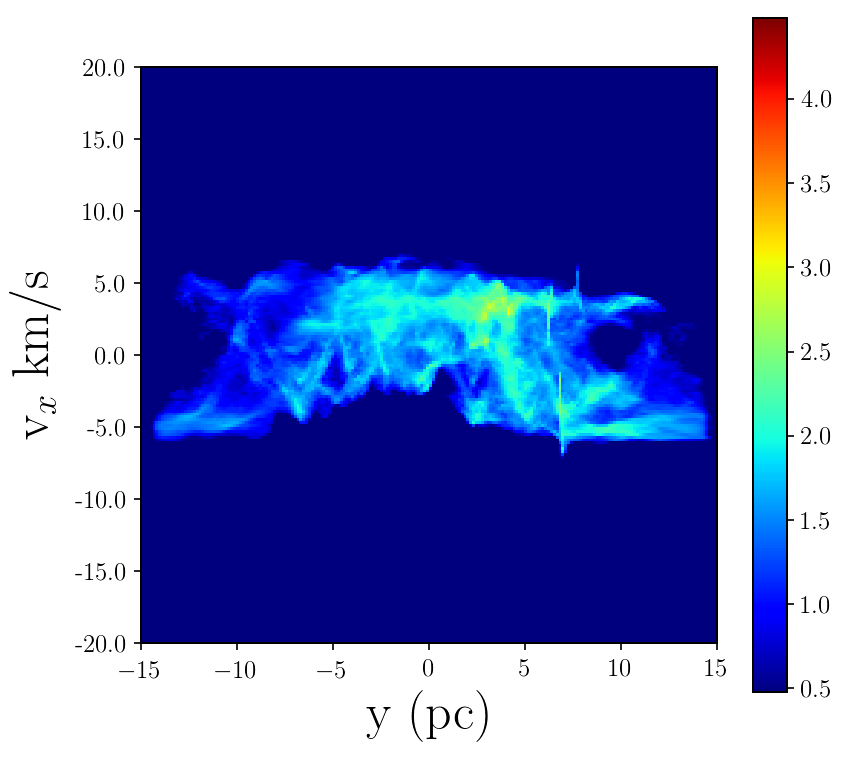}
            \includegraphics[width=0.8\columnwidth]{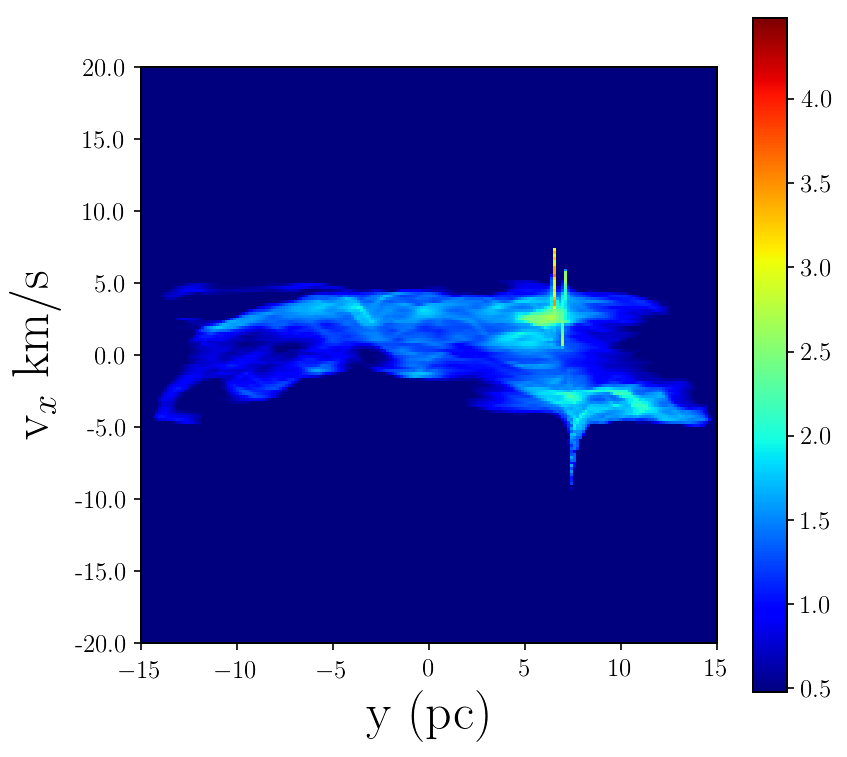}
  \caption{Position-velocity diagrams of Run C simulation. The upper panel is the position-velocity diagram of the clouds prior to the collision. The clouds are separated due to their pre collisional relative velocity. The second plot shows the state of the clouds during collision. The last plot is the position-velocity diagram of the clouds after the collision occurred. The timeline for this position-velocity diagrams are the same as the column density snapshots.}
	\label{fig:runc_pv}
\end{figure}

\begin{figure}
\graphicspath{{Figures/}}
	\centering
	 \includegraphics[width=0.9\columnwidth]{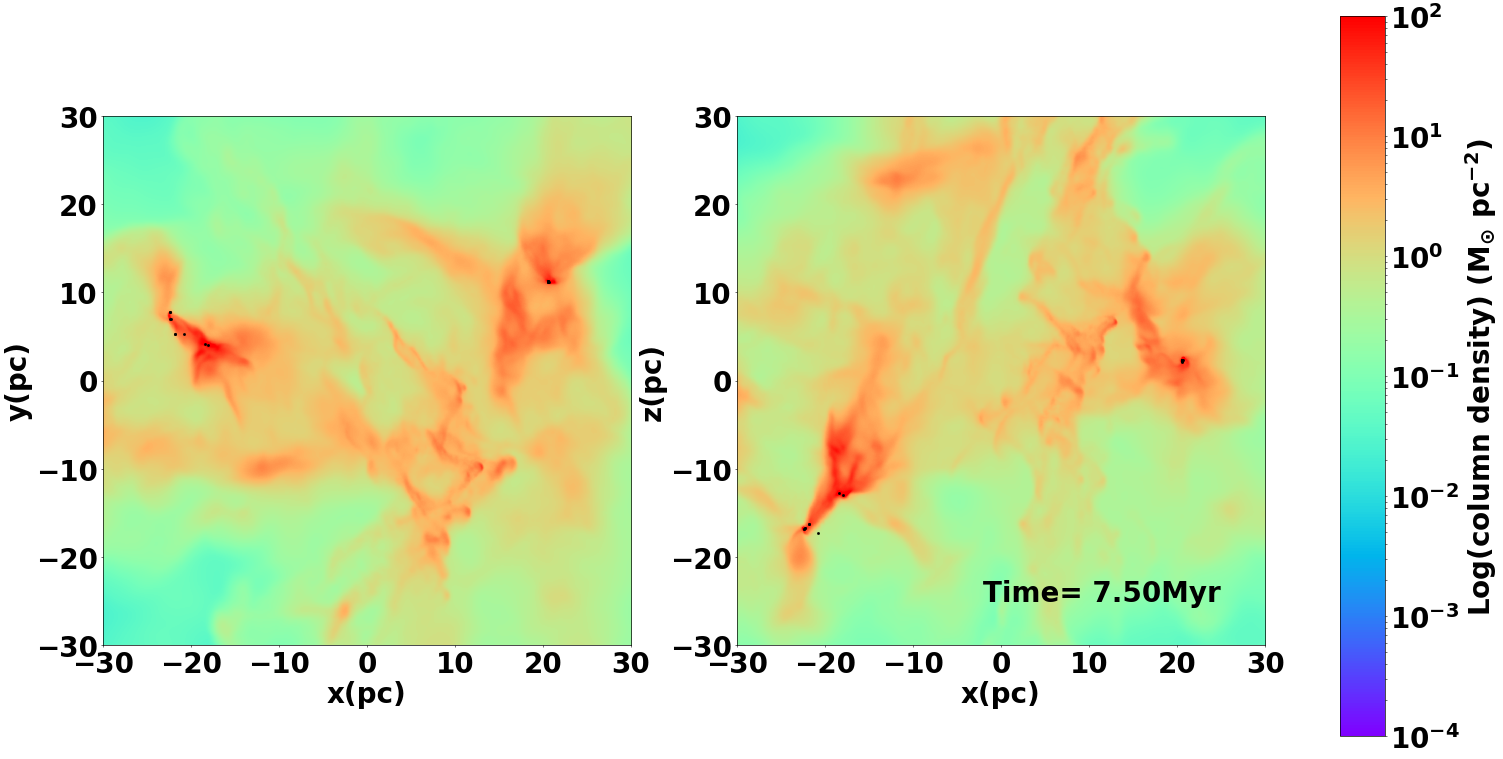}
      \includegraphics[width=0.8\columnwidth]{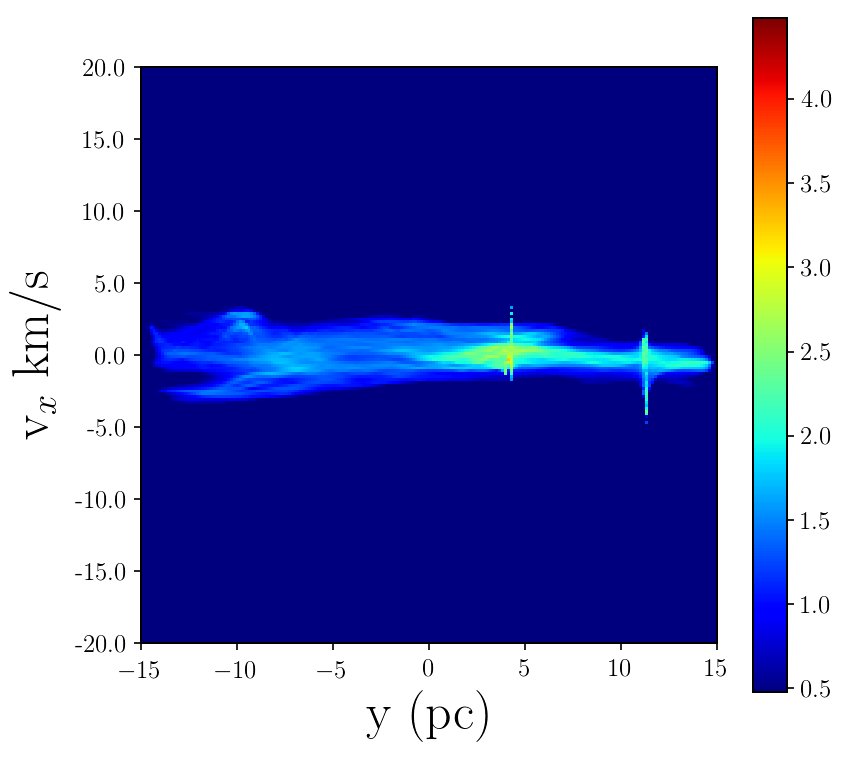}
           
  \caption[column-density]{Column density and position velocity plots of from the Run C simulation snapshots. The first plot shows the column-density at 7.5 Myr of the simulation. The second plot shows the position-velocity diagram of the same run at the same time. From the position-velocity diagram it is hard to distinguish the two clouds since both of them occupy the same space}
     \label{fig:runC_controlled}
	 \end{figure}

\textbf{Sim Run C}\\
In this simulation both the clouds have a virial ratio set at 5.0 which means both clouds are formally unbound and neither would be expected to show vigorous star formation activity in the absence of external factors. This run indeed produces the fewest stars of any run in this paper.\\
\indent Figure \ref{fig:runc_snaps} shows the column density plots of Run C. As before, the clouds meet at around 1.96 Myr, shown in the first panel. The excess turbulent kinetic energy has caused both clouds to expand before striking each other, so that it is already difficult to tell them apart at this early stage.\\
\indent The second panel shows the system at 5.2 Myr. Although there is some dense shocked material originating from the collision, they clouds have largely continued to disperse, indicating that rather little of the initial kinetic energy has been dissipated by the interaction.\\
\indent In the third panel at 7.1 Myr after impact, it is hard to distinguish any features from the original clouds which have survived the collision. There is clearly very little dense gas remaining, and no evidence of the filamentary structures observed in the preceding two runs.\\ 
\indent The position-velocity diagrams from Run C are presented in Figure \ref{fig:runc_pv}. The first panel shows that both clouds are now extended along the velocity axis, and exhibits the  broad bridge feature seen in the other collisional simulations.\\
\indent The second PV image shows that considerable amounts of material, particularly that at negative $y$--locations belonging to the cloud with an initially negative $x$--velocity, has been decelerated by the collision. However this image and the third, which shows the system at 9.0 Myr, shows that several position--velocity features do survive the interaction, even if they are not readily discernible in the PP images. These features remain separated in velocity because the original clouds were unbound, and the collision fails to dissipate enough kinetic energy to bind them together. The third panel also reveals that there is little star--formation activity, and that much of the gas is simply dispersing.\\
\\
Figure \ref{fig:runC_controlled} presents the controlled run of the same simulation. The column-density plot shows the state of the simulation at 7.50 Myr. In this run both the clouds can be identified. It is also quite clear in this run as well that very little dense gas has remained and there is a clear absence of the filamentary structue in this run much like the coliisional run and other controlled runs. From the position-velocity diagram it is once again quite difficult to distinguish each cloud since they both occupy the same width in the velocity axis. From this position-velocity diagram it can be also be determined that most of the gas is just simply dispersing and there is indication of very little star formation occuring at this stage of the simulation. \\
\color{black}
\subsection{Star Formation Rates and Efficiencies}
\begin{figure}
      \centering
      \includegraphics[width=0.8\columnwidth]{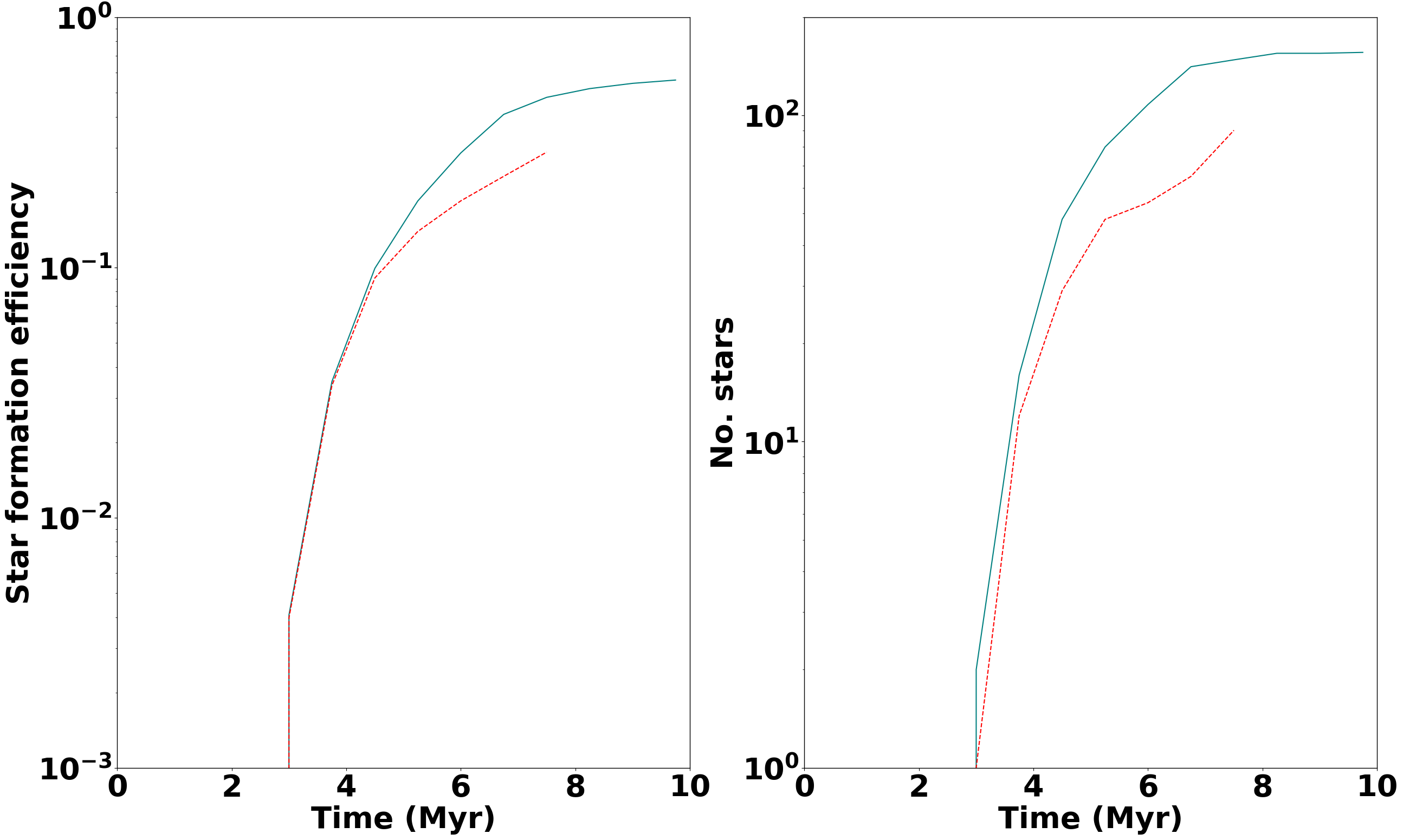}
      \includegraphics[width=0.8\columnwidth]{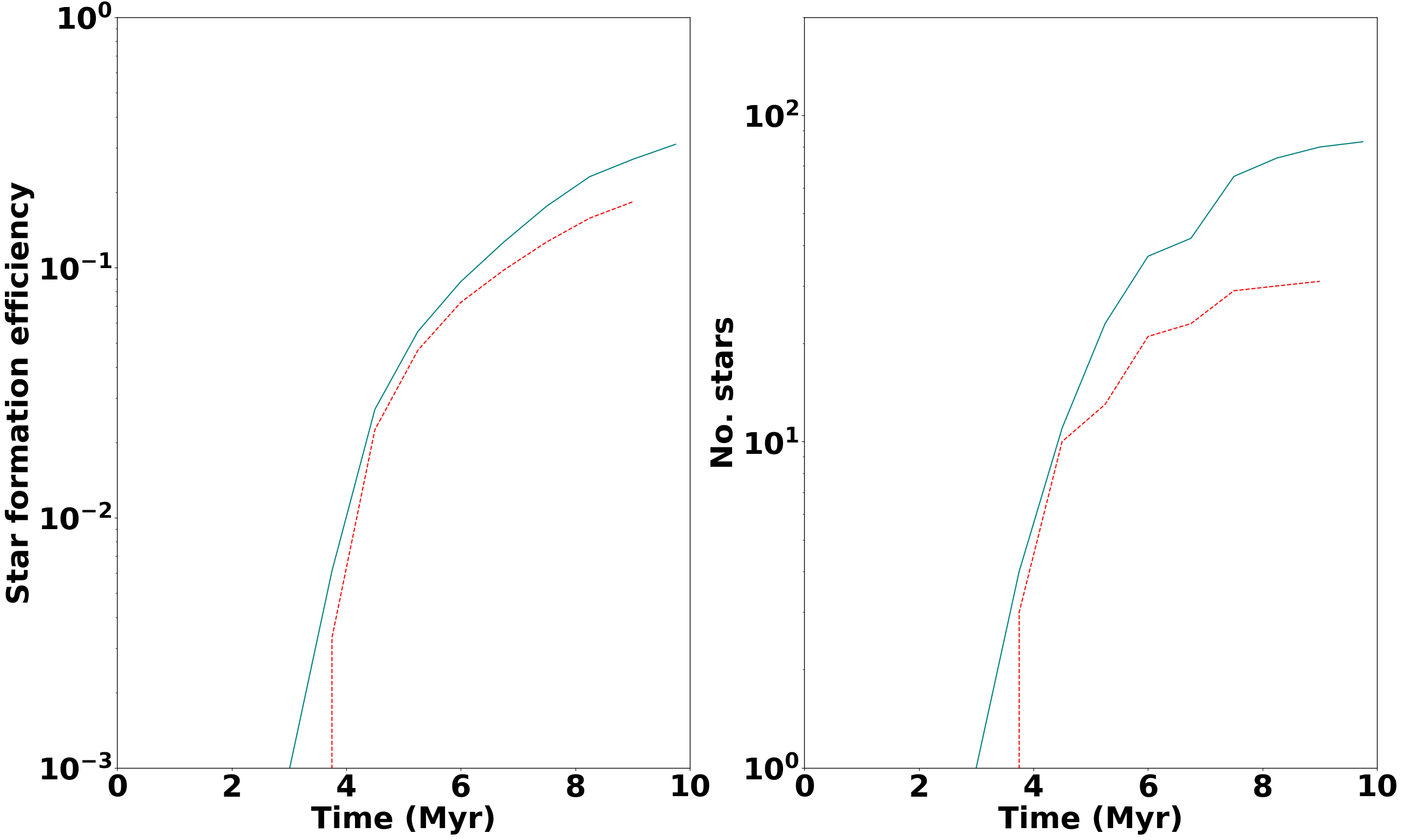}
     \includegraphics[width=0.8\columnwidth]{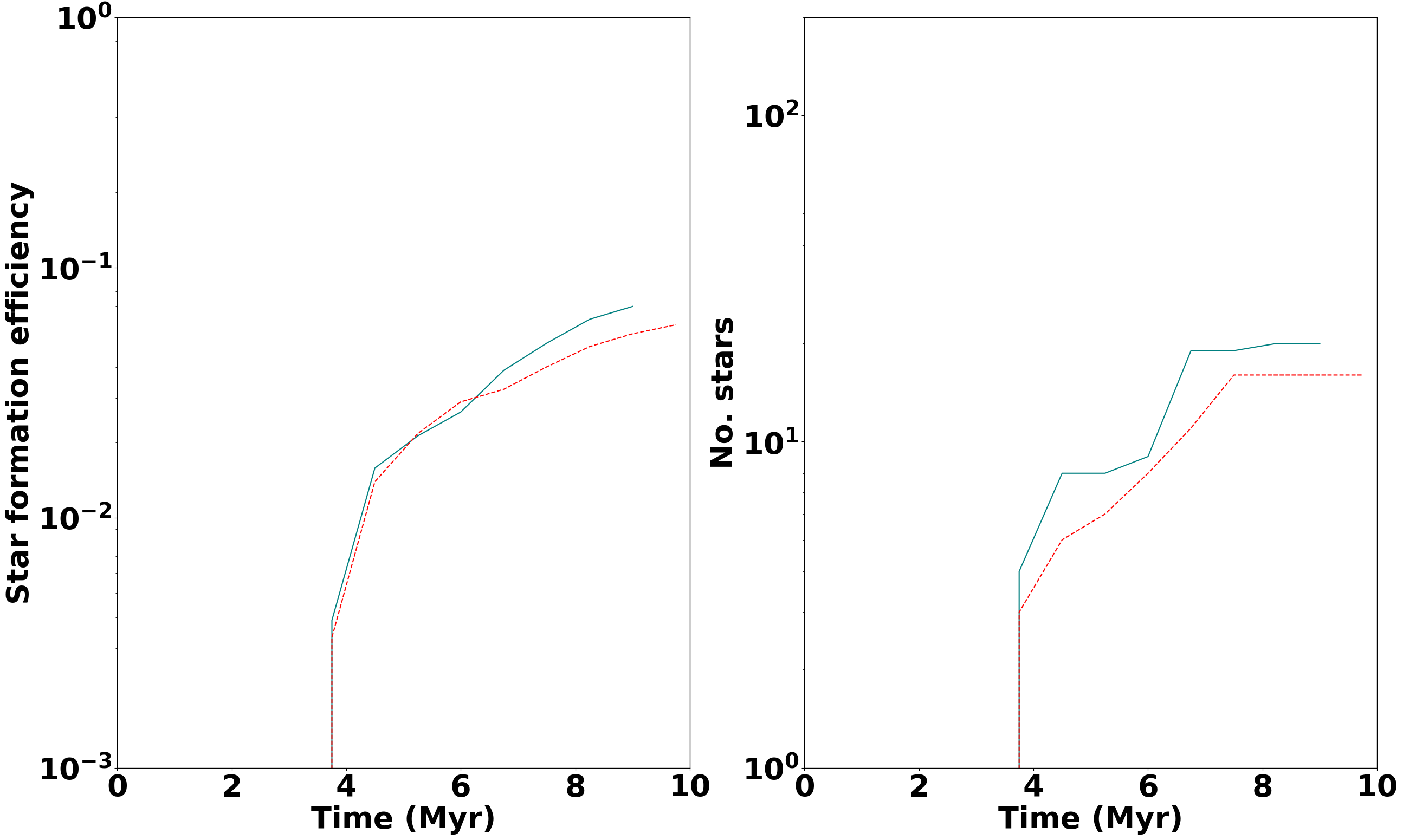}
     \caption{Star formation efficiencies and numbers of stars formed as functions of time. Panel 1 and 2 refer to Run A (the bound--bound collision), panels 3 and 4 to Run B (the bound--unbound collision) and panels 5 and 6 refer to Run C (the unbound--unbound collision). The green lines denotes the collision simulations while the red dotted lines show the star formation efficiency and rate for the control simulations.}
          \label{fig:sfe}
 \end{figure}

Figure \ref{fig:sfe} shows the star formation efficiency and rates for both collision and control simulations. The green lines denotes the collision simulations while the red dotted lines show the star formation efficiency and rate for the control simulations. In all our simulations star formation efficiency is computed from:\\
\begin{equation}
{\rm SFE} = \frac{\rm total\,\,stellar\,\,mass}{\rm total\,\,gas\,\,mass + \rm total\,\,stellar\,\,mass}
 \end{equation}
\indent The first plot shows the star formation efficiency and numbers of stars as functions of time in Run A. The collision simulation produces twice as much stellar mass and twice as many stars as the controlled simulation over the $\approx9$\,Myr span of the simulations. The evolution of the star formation efficiency is initially identical for both simulations scenarios, but they begin to gradually depart from each other at $\sim$ 3.5 Myr. This is about 1.5 Myr after the clouds actually collide, but is roughly the time when the clouds' \textit{centres} pass each other. It is also close to the sum of the collision time and cloud--crushing time. This result implies that the shocking of the bound clouds does indeed  rapidly generate higher density gas compared to the isolated turbulent clouds, increasing the star formation rates and efficiencies, but only modestly.\\
\indent The second plot shows the star formation efficiency and numbers of stars in Run B. Like Run A the star formation rates/efficiencies in the collision and control simulations remain very similar for approximately 2 Myr before beginning to depart from each other. Since there is an unbound cloud present in this run, the star formation rate in the control run is lower than in the corresponding control Run A. However, the effect of the collision on the star formation rate/efficiency is less marked that in Run A, producing only a very modest increase.\\
\indent The third plot shows the star formation efficiency and number of stars as fcuntions of time in Run C. The star formation efficiency is very small in Run C, in both collision and control runs, with the collision having an even smaller effect than in Run B. Therefore, it can be concluded that the effect of collisions on bound clouds is much stronger than it is on unbound clouds. Collisions of this nature are unable to induce unbound clouds to form significantly more stars than they would do if left unmolested.\\ 

\subsection{Boundedness of collision products}
\begin{figure}
      \centering
      \includegraphics[width=1.0\columnwidth]{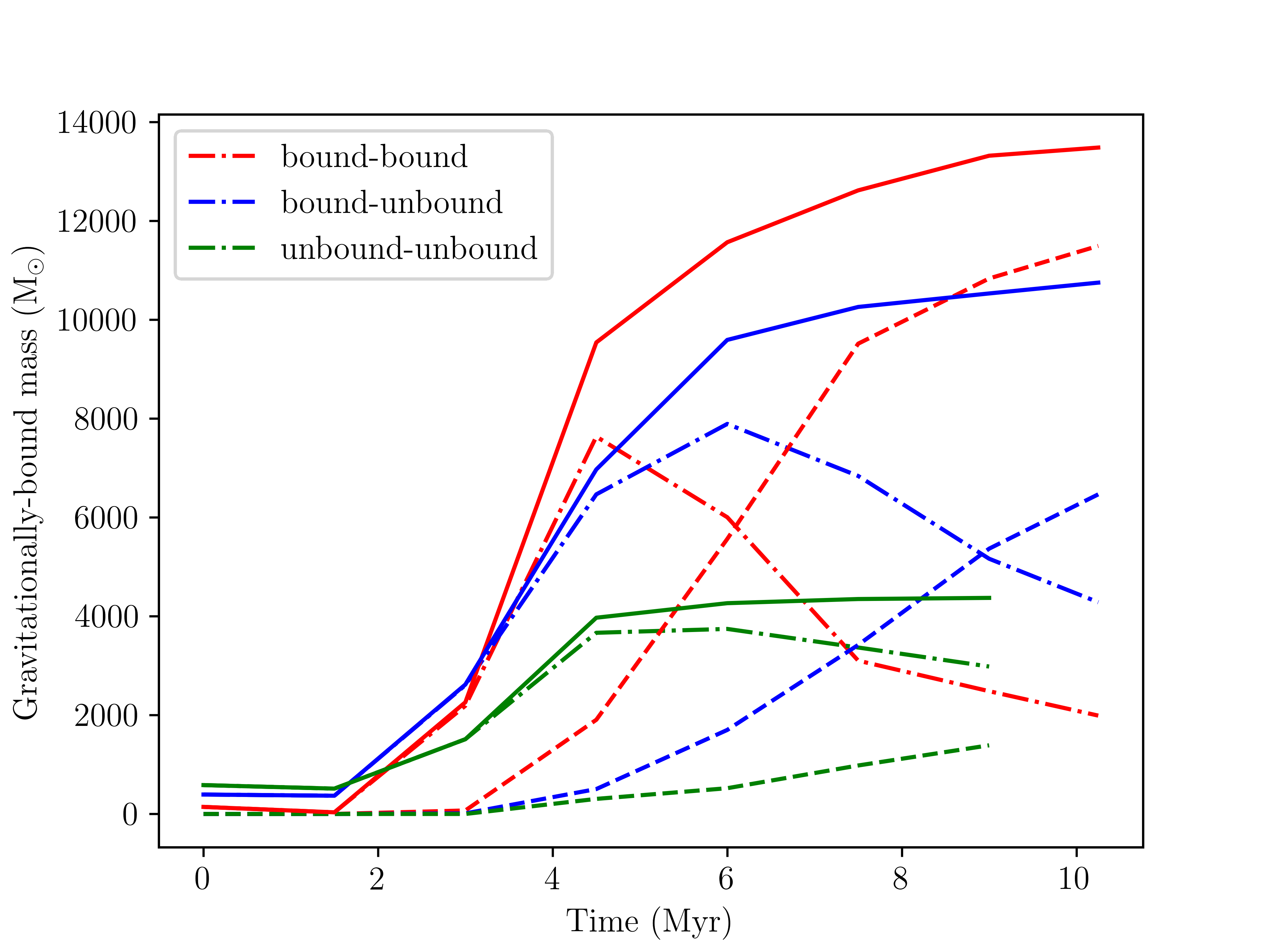}

     \caption{Evolution of the total (solid lines), gas (dashed lines) and stellar (dash--dotted lines) mass bound in the centre--of--mass frame in runs A (red), B (blue) and C (green)}
          \label{fig:bound_mass}
 \end{figure}

\begin{figure}
      \centering
      \includegraphics[width=1.0\columnwidth]{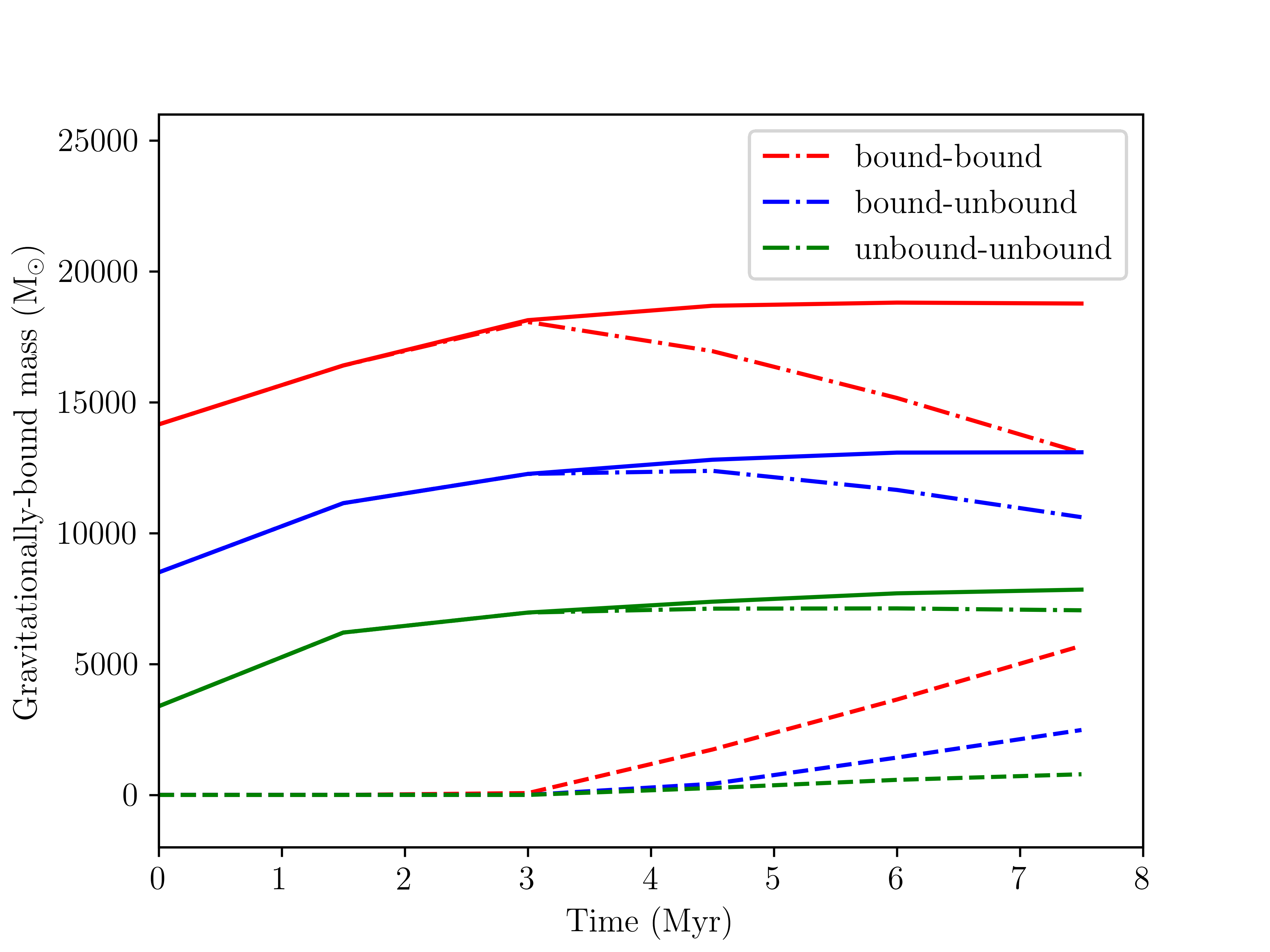}

     \caption{Evolution of the total (solid lines), gas (dashed lines) and stellar (dash--dotted lines) mass bound in the centre--of--mass frame in the control runs A (red), B (blue) and C (green)}
          \label{fig:bound_mass_control}
 \end{figure}

\begin{figure}
      \centering
      \includegraphics[width=0.9\columnwidth]{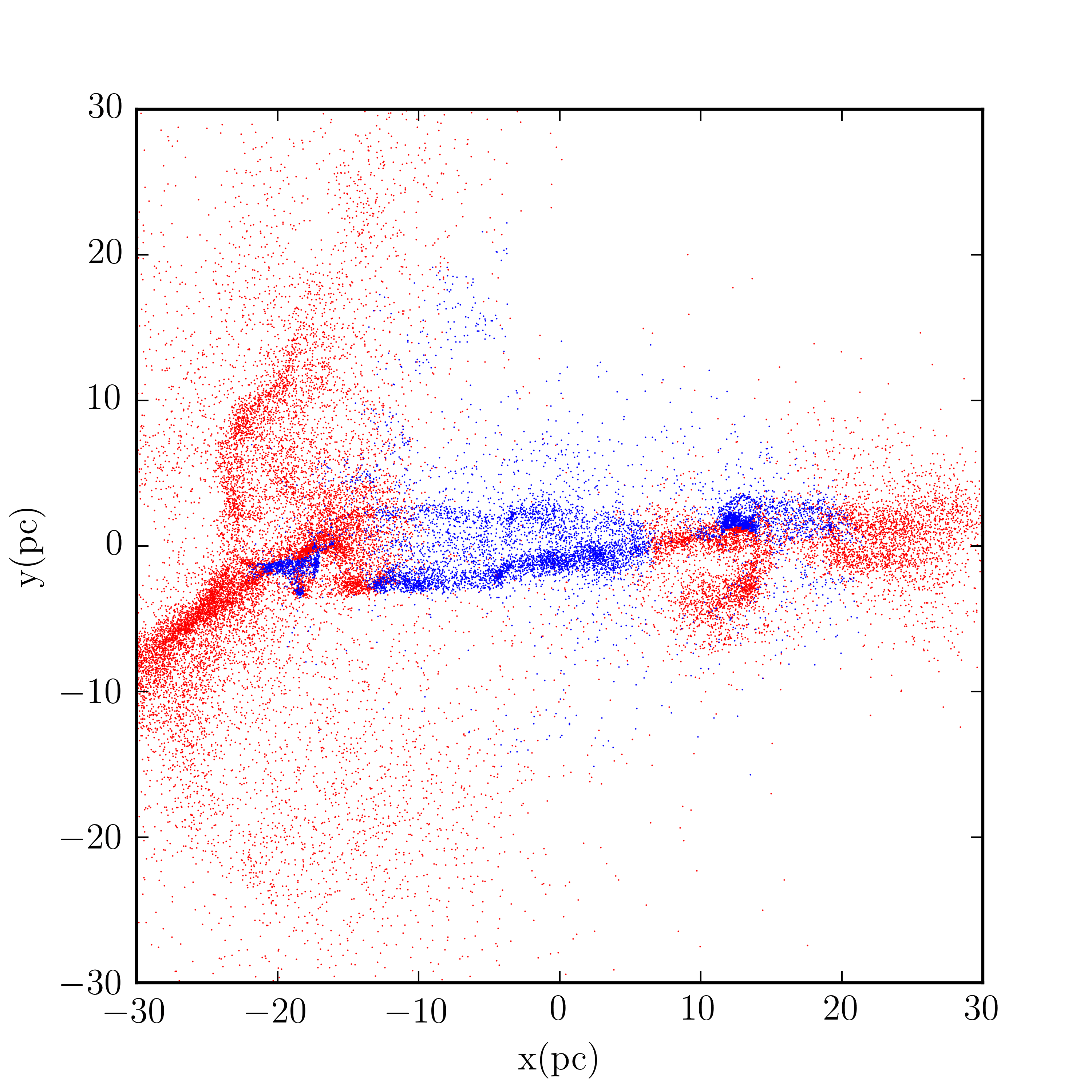}
      \includegraphics[width=0.9\columnwidth]{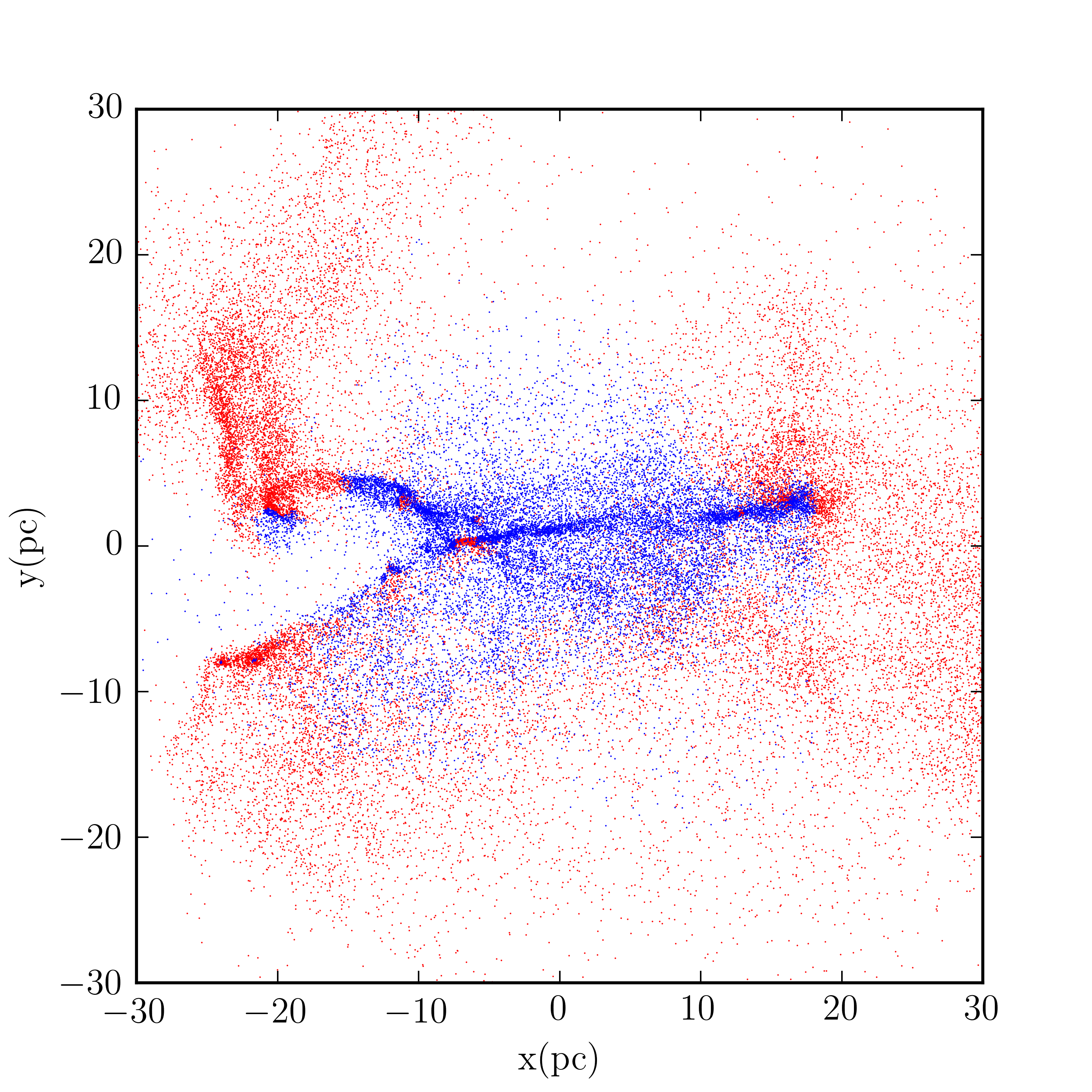}
     \includegraphics[width=0.9\columnwidth]{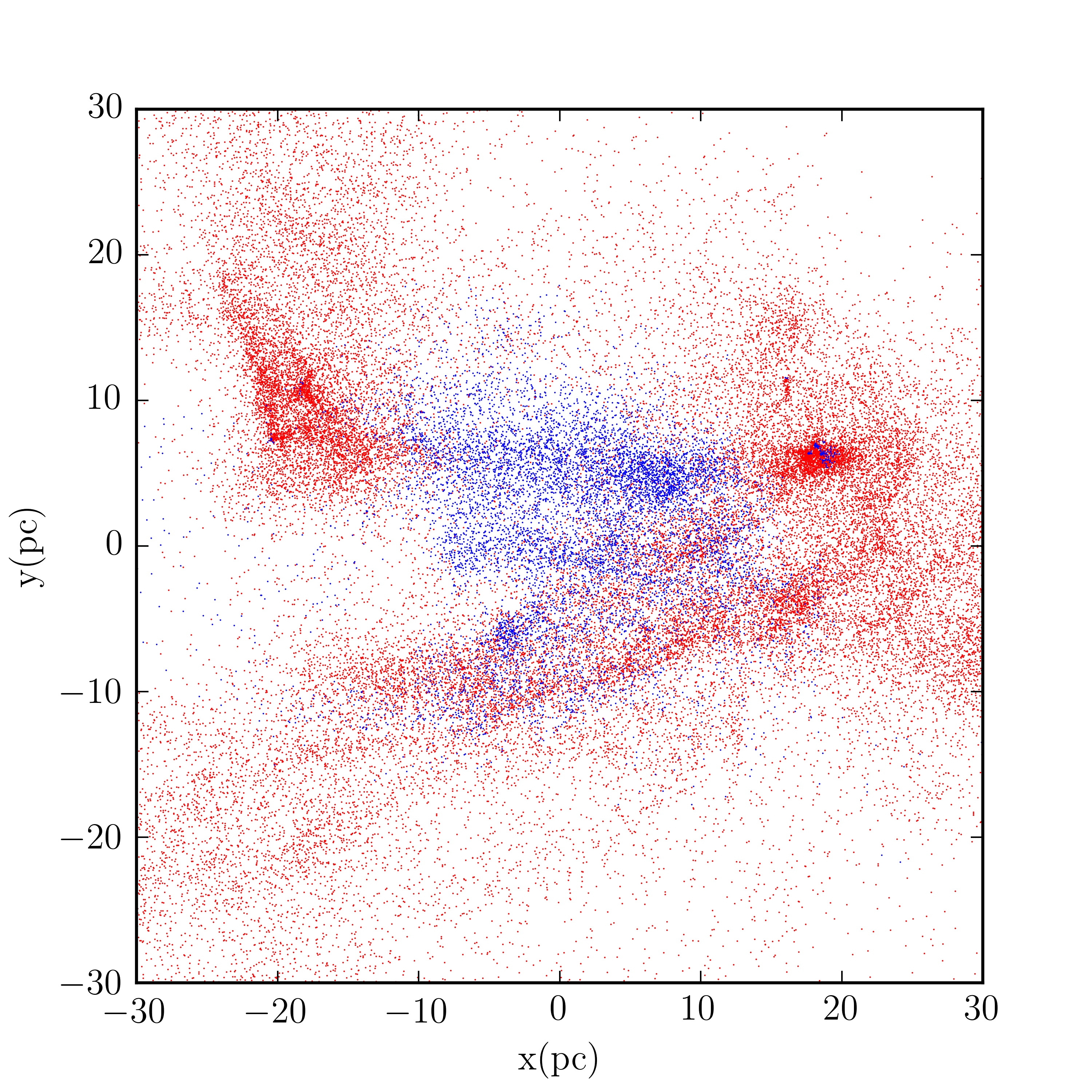}

     \caption{Particle plots of the final states of Runs A (first panel), B (second panel) and C (third panel) with individual gas particles coloured blue if they are bound and red if unbound, in the COM frame.}
          \label{fig:bnd_images}
 \end{figure}

Regardless of whether the individual clouds are bound before the collision, the initial relative velocity and total mass present in all of these simulations is such that the clouds are not initially bound \textit{to each other} in the centre--of--mass (COM) frame of reference.\\
We here examine to what extent the distribution of stars and gas resulting from these collisions results in a single object bound by its own self--gravity in the COM frame. For each gas and sink particle in the simulations, we compute the kinetic energy in the COM frame, the gravitational potential energy due to all the other particles, and the sum of these quantities. Particles for which this sum is negative are regarded as being bound, in the sense that the are expected to form part of a single coherent object whose centre of mass is at rest in the COM frame. We plot the quantities of mass obeying this criterion as a function of time for the three simulated cases in Figure \ref{fig:bound_mass}.\\
It is evident from this plot that the quantity of bound mass in all three simulations is initially negligible, but rises rapidly during the collision as material is decelerated so that its velocity is small in the COM frame and it becomes bound in the potential well developing there. The rate of increase in bound mass is fastest for the bound--bound Run A, and slowest for the unbound--unbound Run C, indicating that matter is more likely to become bound in the COM frame from clouds which are individually more strongly bound. Put another way, the energy dissipation in the collision is sufficient to cause only a small fraction of the material from colliding intrinsically--unbound clouds to become bound in the COM frame.
The quantities of material becoming bound plateau in all cases at around 8\,Myr, and the quantity of bound \textit{gas} declines in all cases as bound material is converted to stars.\\
We conclude that colliding clouds which are themselves intrinsically bound is able to leave a very large fraction of the total mass bound in the COM frame, resulting effectively in a single merged object. However, collisions of intrinsically unbound clouds in this case leave only a small fraction (here, approximately 20\%) of the total mass bound in the COM frame.\\
We compare these results with the evolution of the quantities of bound mass in the control simulations, shown in Figure \ref{fig:bound_mass_control}. The quantities of bound mass increase due to the dissipation of turbulence, but swiftly plateau at considerably larger values than in the corresponding collision simulations, since the clouds in the control simulations have no relative velocity and are thus by definition bound to one another.
In Figure \ref{fig:bnd_images}, we show particle plots of the end--states of runs A, B and C where particles are coded blue if they are bound and red if they are unbound (again, in the COM frame). We see that, in the cases were filaments form, the gas in the filaments is mainly bound, suggesting that the filaments are likely to be long--lived structure.
\color{black}
\section{Discussion and Conclusions}
We have examined the effects of collisions on low--mass turbulent molecular clouds on the grounds that such clouds are the most common and that collisions between them are thus the most common type of such event, since cloud mass functions are negative power laws \citep{2015ARA&A..53..583H}. We choose to collide clouds at a canonical velocity of 10\,km\,s$^{-1}$, which is substantially higher than the escape velocity of our model clouds.\\

\begin{figure}
      \centering
      \includegraphics[width=0.8\columnwidth]{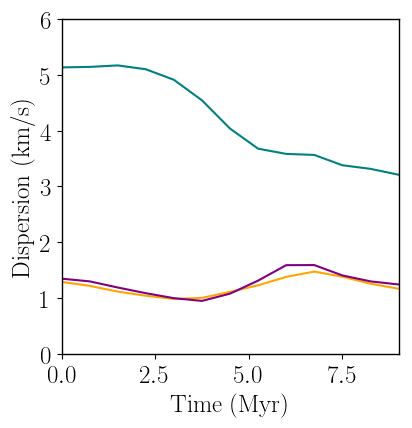}
      \includegraphics[width=0.8\columnwidth]{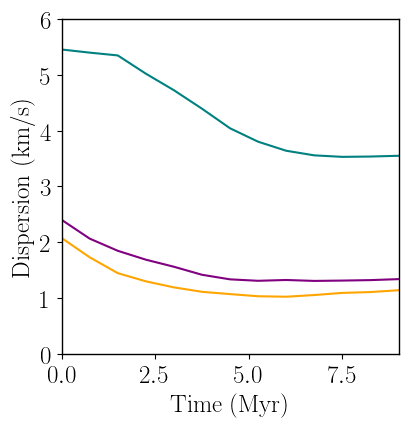}
     \includegraphics[width=0.8\columnwidth]{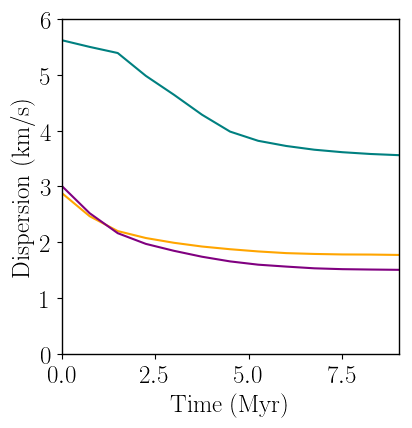}
     \caption{Velocity dispersions in the $x$-- (teal lines) $y$-- (orange lines) and $z$--directions (purple lines) as functions of time in Runs A (first panel), B (second panel) and C (third panel). Dispersions in the $x$--direction include the clouds' relative bulk velocities.}
          \label{fig:sigma}
 \end{figure}

\indent Many simulations on cloud-cloud collision have stopped at around the freefall time or at times when a certain percentage of the gas has been converted into sink particles. However, we have run our simulations to $\sim$ 10 Myr since we intend to understand the evolution of cloud collisions without feedback from massive stars as completely as possible, and we find in particular that the fractions of material left bound by the collisions do not approach settled values until approximately this time. It is also worth pointing out that 2 of our simulations involve unbound clouds for which it is not quite obvious that the freefall time is the most relevant for them. We are at present running a second suite of simulations which involves ionising feedback from massive stars and these set of simulations in this paper will be used as a baseline to understand the effects of feedback in the evolution of cloud collisions.\\
\color{black}
\indent Collisions at velocities substantially larger than their escape velocities can have profound effects on the morphology of turbulent clouds, even though they may already be strongly substructured by turbulence prior to their interaction. The structure created by the turbulence is in fact crucial to understanding the outcome of these encounters. A wealth of earlier work has examined the problem of the collision of two uniform (or at least smooth) and \textit{non--turbulent} clouds \citep[e.g][]{1998ApJ...497..777K,2013MNRAS.431..710M,2015MNRAS.453.2471B}. These simulations produce a thin smooth shocked layer which is prone to several instabilities. The gravitational instability \citep[e.g.][]{1994A&A...290..421W} induces the layer to fragment. Having no support or dispersion perpendicular to the collision axis, the fragments fall in towards it, creating a flattened cluster \citep{2015MNRAS.453.2471B}. The non--linear thin--shell instability may also act to generate structure in the shocked layer \citep{2013MNRAS.431..710M}.\\
\indent Collisions of turbulent clouds have fundamentally different outcomes. The variations in density in the two clouds allow them to interpenetrate each other, so no smooth shocked layer is able to form. A pair of initially spherical clouds of which at least one is bound form filamentary structures visible in both position--position and position--velocity plots, where the filament is approximately parallel to the collision axis. Dense substructure from the original clouds often survives the collision and migrates to the ends of the filaments. These denser regions are usually the main sites of star formation in the collision product. However, if neither cloud is bound, high-velocity collisions fail to generate filaments and the collision remnant continues to disperse.\\
\indent The effect of such collisions on the star formation process in the parent clouds is modest in the case where at least one cloud is bound, and negligible in the case when neither cloud is bound. Collisions of the kind modelled here are thus not efficient in inducing unbound clouds to form stars faster or more efficiently than they would have otherwise. This is despite the fact that the collisions produce significant quantities of dense gas via shocks. The production of dense gas, however, a necessary but not sufficient condition for the formation of stars. If, as is the case here, the velocity dispersion in the gas is too large, the timescale on which condensations are disrupted by fluid flows can be shorter than the local freefall time, preventing gravitational collapse.\\
\indent It does not appear from the results presented above that any of the simulations presented here result in the \textit{complete merger} of the two clouds into a single object. In Runs A, B and C approximately 33\%, 50\% and 80\% of the total mass remains unbound, with the unbound material consisting almost entirely of gas.
In Figure \ref{fig:sigma} we plot the velocity dispersions in the gas along the principal coordinate axes.
Along the $y$-- and $z$--axes, we observe a gentle decline in velocity dispersion as the clouds' turbulent velocity fields dissipate, modulated in the case of Run A by the contraction of the bound clouds.\\
\indent Along the $x$--axis (the collision axis), the velocity dispersions are rather steady until the clouds meet at approximate 2\,Myr. There then follows a period of decline lasting for around the cloud--crushing time (also approximately 2\,Myr), as the clouds enter the shocked region produced by the collision. The decline is due to the dissipation of the clouds' \textit{bulk} kinetic energy by the isothermal shocks generated in the gas by the collision. However, owing to the substructure created in the clouds by turbulence before the collisions, this process is not very efficient (this is discussed in more detail in Paper II where the effects of varying the impact parameter and velocity of the collision are examined). Indeed, in none of these simulations is sufficient bulk kinetic energy lost to render the clouds completely bound.
The velocity dispersions in Figure \ref{fig:sigma} appear to be large enough to render the collision products unbound even in Run A, since they are comfortably in excess of the escape velocity of the original clouds. This is due to the fact that the velocity dispersion is inferred from the surviving gas, and the gas which is not converted stars is most likely to material which is unbound and therefore moving at higher velocity. Observations of the gas at late stages in the simulations therefore automatically pick out material which was not strongly decelerated by the encounter.\\
\color{black}
\indent Magnetic fields, which are neglected in these simulations, are likely to influence cloud--cloud collisions in several ways, and other authors have begun to investigate this complex issue. In particular \cite{2017ApJ...835..137W} use ideal--MHD AMR simulations to investigate the influence of the orientation of an initially--smooth magnetic field relative to the collision axis. They find, not surprisingly, that the further the field is from being parallel to the collision axis, the less high--density gas is generated early in the collision. This is a result of the ability of magnetic fields perpendicular to the collision axis to cushion the impact. This effect may have important implications for the likelihood of the collision product becoming gravitationally bound, since it allows kinetic energy to be converted into magnetic energy (although, of course, this energy could in principle still be yielded back to the gas when the magnetic field relaxes).\\
\indent There is corollary of this, which \cite{2017ApJ...835..137W}'s simulations do not really progress long enough to explore. A field parallel to the collision axis fails to cushion the collision, but would resist the tendency observed in the simulations here involving bound clouds, and those on Paper II, for the collision product to contract towards the collision axis, forming a thick filament. The global and long--term influence of magnetic fields on cloud--cloud collisions thus remains an open question.\\
\section*{Acknowledgements}



\bibliographystyle{mnras}
\bibliography{Bibliography} 








\bsp	
\label{lastpage}
\end{document}